\documentclass{article}
\usepackage{graphicx}
\usepackage[a4paper]{geometry}
\usepackage{comment}
\usepackage{amsmath}
\usepackage{xcolor}
\usepackage{hyperref}
\usepackage{cleveref}

\usepackage[thinc]{esdiff}
\usepackage{mathtools,amssymb}
\usepackage{xcolor}
\usepackage{mathrsfs} 
\usepackage[version=4,arrows=pgf-filled,
textfontname=sffamily,
mathfontname=mathsf]{mhchem} 
\usepackage{subcaption}
\usepackage{overpic}
\usepackage{multirow}
\usepackage{amsmath, amsfonts}
\usepackage{physics}
\usepackage{booktabs}
\usepackage{subcaption}
\renewcommand{\vec}[1]{\boldsymbol{#1}}

\usepackage{listings}
\usepackage[
  style=numeric,
  maxbibnames=2,
  minbibnames=2,
  giveninits=true,
  doi=true,
  url=false,
  isbn=false
]{biblatex}
\DeclareFieldFormat[article]{title}{#1}
\title{\vspace{-1cm}Coupled chemotactic fronts in heterogeneous sensor-consumer\\ cell mixtures}
\author{Marjorie Watts$^{1,*}$, Carles Falc\'o$^{1}$, Giulia L. Celora$^{1,2}$}
\date{}

\begin{document}

\maketitle

\begin{center}
{$^{1}$Mathematical Institute, University of Oxford, OX2 6GG Oxford, United Kingdom}\\
{$^{2}$School of Mathematics, University of Bristol, Fry Building, BS8 1UG Bristol, United Kingdom}

{$^*$Correspondence: marjorie.watts@maths.ox.ac.uk
}
\end{center}

\begin{abstract}
   Chemotaxis underlies the collective migration of cell populations in developmental processes and immune responses. While the theoretical investigation of single-cell-type collective chemotaxis has received considerable attention, heterogeneous chemotaxis involving multiple interacting cell types remains poorly understood. Here, we generalise a model of heterogeneous self-generated chemotaxis and analyse the resulting collective migration patterns. We show that coupled migration between two cell types gives rise to \emph{propagating terraces}---coupled travelling fronts moving at different speeds. While a sensor-only population leads the migrating collective, a slower mixed sensor–consumer population follows. Our analysis reveals that these fronts are coupled via the dynamics of the self-generated chemoattractant gradients. We derive analytical expressions for the migration speeds of the two fronts just in term of model parameters and experimentally measurable quantities. Our analytical results reveal that heterogeneity can enhance long-range migration via self-generated chemotaxis for sensor cells. While sensor cells can leverage benefit from mixing with consumer cells, the latter migrate more efficiently when mixing with cells of the same type. Together, our results provide a comprehensive theoretical framework for understanding heterogeneous self-generated chemotaxis.
\end{abstract}

\section*{Introduction}
Cell migration is a fundamental biological process that underlies embryonic development, tissue homeostasis, wound healing, and immune surveillance~\cite{Alanko2023,McLennan2015,Ratnayake2021}, while its dysregulation drives pathological conditions such as cancer metastasis~\cite{Clark2015, Qu2023, Yamamoto2023}, the leading cause of cancer-related mortality.
Cell migration is governed by a complex interplay of cell-intrinsic and cell-extrinsic factors which span across different spatial and temporal scales. A prominent example of a cell-extrinsic cue is \emph{chemotaxis}: the directed movement of a cell or organism in response to chemical gradients in its environment, which allows cells to navigate toward favourable conditions or away from harmful stimuli~\cite{Hillen2008}. 
Chemotaxis has traditionally been understood in terms of deterministic migration along pre-existing external gradients, despite limited evidence \emph{in vivo}. 
Recent experiments instead highlight locally self-generated signals, where cells actively shape their environment to guide collective migration~\cite{Alanko2023,dona2013directional, stock2022self, Ucar2025}.

Many continuum models for self-generated chemotaxis build on the seminal work of Keller and Segel~\cite{Keller1971}, who developed a mathematical model to describe the travelling bands of \emph{E. coli} observed experimentally when cells were placed at one end of a capillary tube containing oxygen and a nutrient source.
Their phenomenological description captured the coupled evolution of cell density and substrate concentration, incorporating both random (diffusive) and directed (chemotactic) cell migration alongside local substrate consumption.
Extensions of this framework have explained directed cell motions in a range of biological systems~\cite{Celora2026-2,bhattacharjee2021chemotactic,ferguson_statistical_2017,Ucar2025,SaragostiEtAl2010}, including bacterial colonies, immune cell populations, and cancer. Chemotaxis models based on the work of Keller and Segel have also inspired a substantial body of analytical work on the existence and structural properties of its solutions (\emph{e.g.},\cite{blanchet2008infinite, calvez2006volume, CalvezCorrias2008,Hillen2008}), and theoretical investigations on stability of chemotactic fronts~\cite{Alert2022} and their travelling speed~\cite{Narla2021}.
For instance, Narla et al.~\cite{Narla2021} elucidated the relation between measurable molecular and environmental parameters and the migration properties of chemotacting growing populations by deriving approximate analytical expression for the travelling wave solutions in a variation of the Keller--Segel model that includes cell proliferation. 
While the majority of Keller–Segel-type models assign uniform chemotactic and sensing characteristics to all cells in the population, however, biological tissues are rarely so simple. 
Evolutionary pressures often drive cells to specialise for distinct tasks, resulting in heterogeneity within cell populations~\cite{p2025phenotypic,salek2019bacterial}. 
The consequences of such intra-population heterogeneity for collective self-generated chemotaxis remain poorly understood.

There has been a growing theoretical interest in modelling and understanding self-generated chemotaxis of heterogeneous populations~\cite{macfarlane_impact_2022,lorenzi_phenotype_2025,freingruber_trait-structured_2025,Mattingly2022}. For example, Mattingly and Emonet~\cite{Mattingly2022} used a multi-species Keller--Segel model to show that individuals are spatially organised within migrating cell collectives based on their chemotactic ability. This causes low-performing cells to be gradually lost from the migrating front allowing the population to adapt its phenotypic composition to the environment it traverses, without any gene regulation or mutation. Central to this mechanism is a balance between cell growth continuously regenerating phenotypic diversity and collective migration selectively filtering out under performers. This theoretical prediction was subsequently confirmed experimentally by Vo et al.~\cite{Vo2025}, who directly observed that migrating \emph{E. coli} populations became rapidly and reversibly enriched in high-performing chemotactic phenotypes. Heterogeneity is also leveraged in physiological responses, such as during immune response where the different types of immune cells cooperate to collectively migrate towards its target. For example, U\c{c}ar et al.~\cite{Ucar2025} have shown that the co-migration of T and dendritic (D) cells can be described via a variation of the Keller–Segel model for self-generated chemotaxis that accounts for two distinct cell types: \emph{consumer} (DCs) and \emph{sensor} (TCs) cells. 
While the consumer cells can both generate and follow the chemotactic gradients, sensor cells cannot locally deplete the chemoattractant and reshape its gradient. Consequently, the sensor population can surf along the gradient generated by the consumer cell and position themselves at the leading front---creating a spatially-organised migrating front. 
These examples make clear that phenotypic heterogeneity is a key aspect of collective chemotactic dynamics allowing for spatio-temporal adaptation of migrating cell populations. Yet, 
a comprehensive analytical understanding of self-generating chemotaxis in heterogeneous population is still lacking. 

Here, we aim to partially address this gap, building on recent works on sensor/consumer asymmetry during collective chemotaxis~\cite{Celora2026-2,Ucar2025}. In the original sensor-consumer model proposed by~\cite{Ucar2025}, sensor cells do not consume the chemoattractant. Under this assumption, the model predicts that the dynamics eventually converge to a travelling wave solution, whose speed~\cite{Ucar2025} and spatial structure~\cite{Celora2026-2} have been analytically characterised. 
By allowing sensor cells to also shape the chemoattractant profile, we show that qualitatively different types of migration dynamics are possible. Once both populations are able to migrate in isolation, the system supports propagating terrace solutions consisting of two travelling waves coupled through the shared chemoattractant field. These observations motivate the analytical framework developed below to characterise the coupled travelling waves and their dependence on model parameters. Our analysis reveals an emergent asymmetry between sensor and consumer cells: sensor cells benefit from coupling with consumer cells, whereas consumer cells generally migrate more efficiently when mixed with cells of the same type.

The paper is organised as follows. 
In Section~\ref{sec:model}, we present a minimal model for the self-generated chemotactic migration of a mixed sensor-consumer population.~To establish a baseline for comparison, in Section~\ref{sec:homo}, we first analyse the model predictions in the simplified scenario in which the two cell types are identical, \emph{i.e.}, an homogeneous population. 
In this regime, we derive an explicit expression for the travelling wave speed in terms of the model parameters, allowing us to determine how the invasion behaviour varies across parameter space.
In Section~\ref{sec:heterogeneous chemotaxis}, we consider to the full heterogeneous model and investigate the impact of heterogeneity on collective migration patterns. 
Numerical simulations reveal that migration occurs in the form of two coupled travelling wave fronts that invade at different speeds, which we characterise analytically. 
This analysis leads to a system of coupled non-linear algebraic equations that implicitly define the travelling speeds of the two fronts in terms of model parameters. 
Solving this system numerically, we study the coupled invasion dynamics of heterogeneous sensor/consumer systems for a wide range of parameter values and find that co-migration consistently enables faster propagation of sensor cells than they would do in isolation. 
We conclude by summarising our key findings and outlining future research directions.

\section{A model of heterogeneous, self-generated chemotaxis}\label{sec:model}
\begin{figure}
    \centering
    \begin{subfigure}{0.0\textwidth}
        \captionlistentry{}
        \label{fig:figure1A}
    \end{subfigure}
        \begin{subfigure}{0.0\textwidth}
        \captionlistentry{}
        \label{fig:figure1B}
    \end{subfigure}
        \begin{subfigure}{0.0\textwidth}
        \captionlistentry{}
        \label{fig:figure1C}
    \end{subfigure}
                \includegraphics[width=\linewidth]{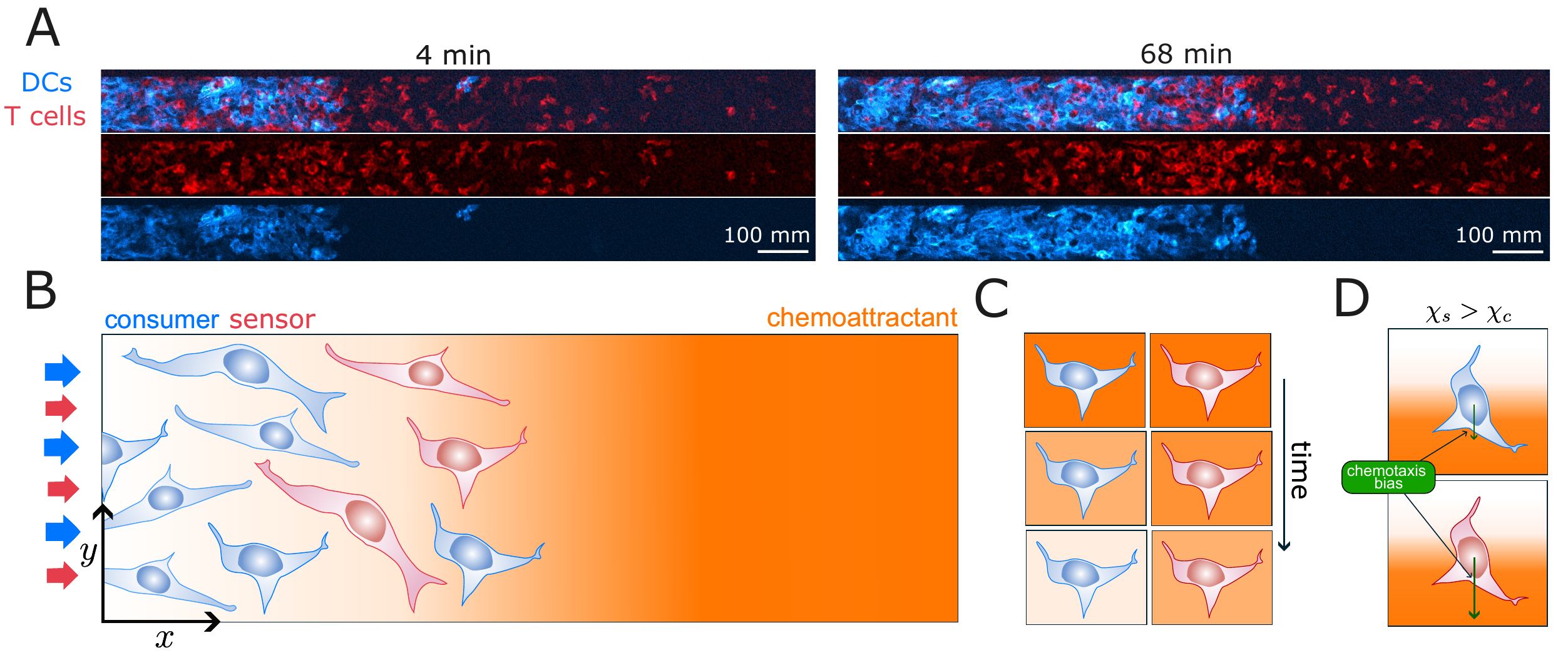}
    \caption{ (A) Microscopy images from microfluidic experiments with labelled DCs (blue) and T cells (red) at two different time points. (Obtained from~\cite{Ucar2025} with permission, licensed under CC BY 4.0) (B) Geometry of the model setup is inspired by the experiments shown in~\Cref{fig:figure1A}: consumer (blue) and sensor (red) cells are initially located at $x=0$ and migrate toward increasing $x$, with the cell and chemoattractant concentrations assumed uniform along the $y$-direction. (C) Temporal dynamics of chemoattractant depletion, with consumer cells (blue) consuming chemoattractant at a higher rate than sensor cells (red). (D) Components of cell motion: random diffusion and directed chemotaxis, with sensor cells (red) exhibiting a stronger chemotactic response than consumer cells (blue).}
    \label{fig:hetero_schematic}
\end{figure}

In this paper, we investigate the coupled migration of two cell types, sensor and consumer cells. We assume that both cell types have the ability for long-range migration via self-generated chemotaxis in response to gradients of the same chemoattractant.
As illustrated in Figure~\ref{fig:hetero_schematic}, we consider a population of cells migrating toward a chemoattractant source located at $x\gg 1$, modulating the chemoattractant concentration and sensing the local gradient it generates.  
We assume that these processes occur on a much faster timescale than cell proliferation, which we neglect.  
The geometry is inspired by the experimental setups considered by U\c{c}ar et al. and Keller--Segel, in which chemoattractant-sensing cells are positioned at one end of a capillary tube filled with a medium that supports motility --see Figure~\ref{fig:figure1A}. 
Under these conditions, the system can be reduced to a one-dimensional model with no dependence on the $y$- or $z$-coordinates.
We denote the concentration of the cells by $\rho_c(x, t)$ and $\rho_s(x, t)$ for the consumer and sensor cell types, respectively, and the concentration of the chemoattractant $a(x, t)$, with units $\text{cells} \,\, \text{mm}^{-1}$ and $\text{mmol} \,\, \text{mm}^{-1}$, respectively.
We assume that sensor and consumer cells differ through irreversible specialisation---\emph{i.e.}, we neglect the possibility of phenotypic switching~\cite{crossley_phenotypic_2024}.
The consumer cells are defined as having an increased ability to consume the chemoattractant, while the sensor cells are defined as having an increased ability to sense the chemical gradient that they are both following.

Under the given assumptions, the chemotaxis-driven migration of a mixed population of sensor/consumer cells is described by the following system of coupled non-linear partial differential equations:
\begin{subequations}\label{eq:dim_sys}
    \begin{align}
        \pdv{\rho_c}{t} &= D_c \pdv[2]{\rho_c}{x} - \kappa\chi \diffp{}{x} \left[ \rho_c \, \diffp{}{x} \Big[ \log(a) \Big]  \vphantom{\diffp{\rho}{x}} \right] ,\label{eq:dim_sys_a}\\
        \pdv{\rho_s}{t} &= D_s \pdv[2]{\rho_s}{x} - (1-\kappa)\chi \diffp{}{x} \left[ \rho_s \, \diffp{}{x} \Big[ \log(a) \Big]  \vphantom{\diffp{\rho}{x}} \right] ,\label{eq:dim_sys_b}\\
        \pdv{a}{t} &= D_a \pdv[2]{a}{x} - r \Big[ \omega \rho_s + (1-\omega)\rho_c \Big] a ,
    \end{align}
\end{subequations}
where constants $D_i$ ($i=c,s,a$) characterise the diffusion of the consumer and sensor cells, and chemoattractant, respectively, while $r$ and $\chi$ denote the total cellular consumption rate and chemotactic sensing abilities of the population.  
The parameters $\kappa$ and $\omega$ determine how the total sensing and consumption capacities, respectively, are distributed between the two cell types. To guarantee that both sensor and consumer cells retain the ability to detect and shape the chemical gradient, while assigning a greater sensing sensitivity to the sensor cells and greater consumption ability to the consumer cells, we assume $0<\omega,\kappa<1/2$.
Setting $\kappa,\,\omega =1/2$, we recover the homogeneous population case, in which both cell types are equally sensitive to the chemoattractant and deplete it at the same rate.
The model defined by U\c{c}ar et al.~\cite{Ucar2025} considers the special case $\omega=0$, when sensor cells are incapable of self-generated chemotaxis, and the sensing abilities of the two cell types are decoupled from $\kappa$ and instead governed by independent parameters $\chi_c$ and $\chi_s$.

Several functional forms for the term representing advection due to chemotactic sensitivity (second term of the RHS of Equations~\eqref{eq:dim_sys_a}--\eqref{eq:dim_sys_b}) have been proposed to simulate different biological mechanisms. 
Two widely used models are the receptor and logarithmic models~\cite{Hillen2008}, both of which describe a monotonic decrease in chemotactic sensitivity with increasing attractant concentration, reflecting receptor saturation. 
The receptor model is based on Michaelis--Menten receptor kinetics, whereas the logarithmic model follows the Weber--Fechner law and is consistent with observations that \emph{E. coli} senses gradients in the logarithm of ligand concentration over a wide range of background concentrations~\cite{Narla2021}. 
We therefore adopt the logarithmic form for the chemotactic sensing function in Equation~\eqref{eq:dim_sys} for its biological relevance and mathematical tractability.

Following~\cite{Ucar2025}, at the inlet ($x=0$), we allow cells to enter the domain by at a given rate $\gamma_{c,s}\geq 0$, while we consider a system with no chemoattractant influx:
\begin{subequations}\label{eq:BCs_dim}
    \begin{align}\label{eq:BCs_inlet}
    \left. - D_c \diffp{\rho_c}{x} + \kappa\chi \rho_c \diffp{}{x} \left( \log a \right)   \right|_{x=0} = \gamma_c, \left. \quad - D_s \diffp{\rho_c}{x} + (1-\kappa)\chi \rho_s \diffp{}{x} \left( \log a \right)  \right|_{x=0} = \gamma_s,\quad \left. D_a \pdv{a}{x} \right|_{x=0} = 0.
\end{align}
Far from the origin, we assume that the cell densities decay to zero, while the chemoattractant reaches its maximal concentration $\alpha$
\begin{equation}\label{eq:BCs_inf}
    \lim_{x \to \infty} \rho_c = \lim_{x \to \infty} \rho_s=0, \quad \lim_{x \to \infty} a = \alpha . 
\end{equation}
\end{subequations}

which ensures that no cells enter or leave the domain through this boundary. Hence, in the absence of cell growth or death, Equations~\eqref{eq:BCs_inlet}--\eqref{eq:BCs_inf} imply that $\gamma_c$ and $\gamma_s$ completely determine the total cell mass in the system, providing additional constraints that we exploit in the analysis below.

\subsection{Non-dimensional model} Since we are interested in studying the general model behaviour, we non-dimensionalise the system~\eqref{eq:dim_sys}--\eqref{eq:BCs_dim} following the scalings detailed in Appendix~\ref{app:non_dim} to obtain
\begin{subequations}\label{eq:nondim_sys}
    \begin{align}
        \pdv{\rho_c}{t} &= \tilde D_c \pdv[2]{\rho_c}{x} - \kappa\tilde\chi \diffp{}{x} \left[ \rho_c \, \diffp{}{x} \left( \log a \right)   \vphantom{\diffp{\rho}{x}} \right] , \label{eq:nondim_sys_a}\\
        \pdv{\rho_s}{t} &= \tilde D_s \pdv[2]{\rho_s}{x} - (1-\kappa)\tilde\chi \diffp{}{x} \left[ \rho_s \, \diffp{}{x} \left( \log a \right)  \vphantom{\diffp{\rho}{x}} \right] , \label{eq:nondim_sys_b}\\
        \pdv{a}{t} &= \pdv[2]{a}{x} - \Big[ \omega \rho_s + (1-\omega)\rho_c \Big] a \label{eq:nondim_sys_c},
    \end{align}
\end{subequations}
with similarly rescaled boundary conditions
\begin{align}
    \left. \pdv{a}{x} \right|_{x=0} = 0 \quad &\text{and} \quad \lim_{x \to \infty} a = 1 ,\label{eq:nondim_BCs_a}\\
    \tilde D_i \left. \pdv{\rho_i}{x} \right|_{x=0} = \tilde\gamma_i \quad &\text{and} \quad \lim_{x \to \infty} \rho_i = 0, \qquad i=c,s \, .\label{eq:nondim_BCs_rho}
\end{align}
Note that in writing~\eqref{eq:nondim_BCs_rho}, we have used the Neumann conditions applied to $a$ to simplify the expression of the cell fluxes.
Equations~\eqref{eq:nondim_sys}--\eqref{eq:nondim_BCs_rho} are closed by imposing initial conditions. 
Following~\cite{Ucar2025}, we here consider a constant initial chemoattractant profile, while the cell density profiles of both cell types follow a sigmoidal curve so that they decrease to zero as $x\to\infty$ (see Appendix~\ref{app:ICs} for more details).

All numerical simulations are performed using a finite volume scheme to ensure conservation of mass \cite{carrillo2015finite}; further details can be found in Appendix~\ref{app:finite_volume}.
Parameter values for the simulations have been adapted from \cite{Ucar2025}, and are presented in Table~\ref{tab:params}.

\section{Migration of homogeneous cell populations}\label{sec:homo}
We start by considering collective chemotaxis of homogeneous cell populations. 
This allows us to summarise existing results on travelling wave solutions for this system from the literature~\cite{Keller1971,Ucar2025} and to establish a baseline against which comparing the migration of heterogeneous sensor/consumer populations.
Setting $\rho \coloneq \rho_s+\rho_c$, $D_\rho\coloneq D_s = D_c$, $\kappa = \omega = 1/2$, and $\tilde\gamma \coloneq \tilde\gamma_s + \tilde\gamma_c$, system~\eqref{eq:nondim_sys} reduces to
\begin{subequations}\label{eq:nondim_sys_homo}
    \begin{align}
        \pdv{\rho}{t} &= \tilde D_\rho \pdv[2]{\rho}{x} - \tilde\chi \diffp{}{x} \left[ \rho \, \diffp{}{x} \left( \log a \right) \vphantom{\diffp{\rho}{x}} \right], \label{eq:nondim_sys_homo_a} \\
        \pdv{a}{t} &= \pdv[2]{a}{x} - \rho\, a, \label{eq:nondim_sys_homo_b}
    \end{align}
with boundary conditions
\begin{equation}
    \left. \pdv{a}{x} \right|_{x=0} = 0 \quad \text{and} \quad \lim_{x \to \infty} \pdv{a}{x} = 0 ,\quad 
    \tilde D_\rho \left. \pdv{\rho}{x} \right|_{x=0} = \tilde\gamma \quad \text{and} \quad \lim_{x \to \infty} \rho = 0 .
\end{equation}
\end{subequations}

\begin{figure}[htb]
    \centering
    \begin{subfigure}{0.0\textwidth}
        \captionlistentry{}
        \label{fig:figure2A}
    \end{subfigure}
        \begin{subfigure}{0.0\textwidth}
        \captionlistentry{}
        \label{fig:figure2B}
    \end{subfigure}
        \begin{subfigure}{0.0\textwidth}
        \captionlistentry{}
        \label{fig:figure2C}
    \end{subfigure}
            \begin{subfigure}{0.0\textwidth}
        \captionlistentry{}
        \label{fig:figure2D}
    \end{subfigure}
    \includegraphics[width=\linewidth]{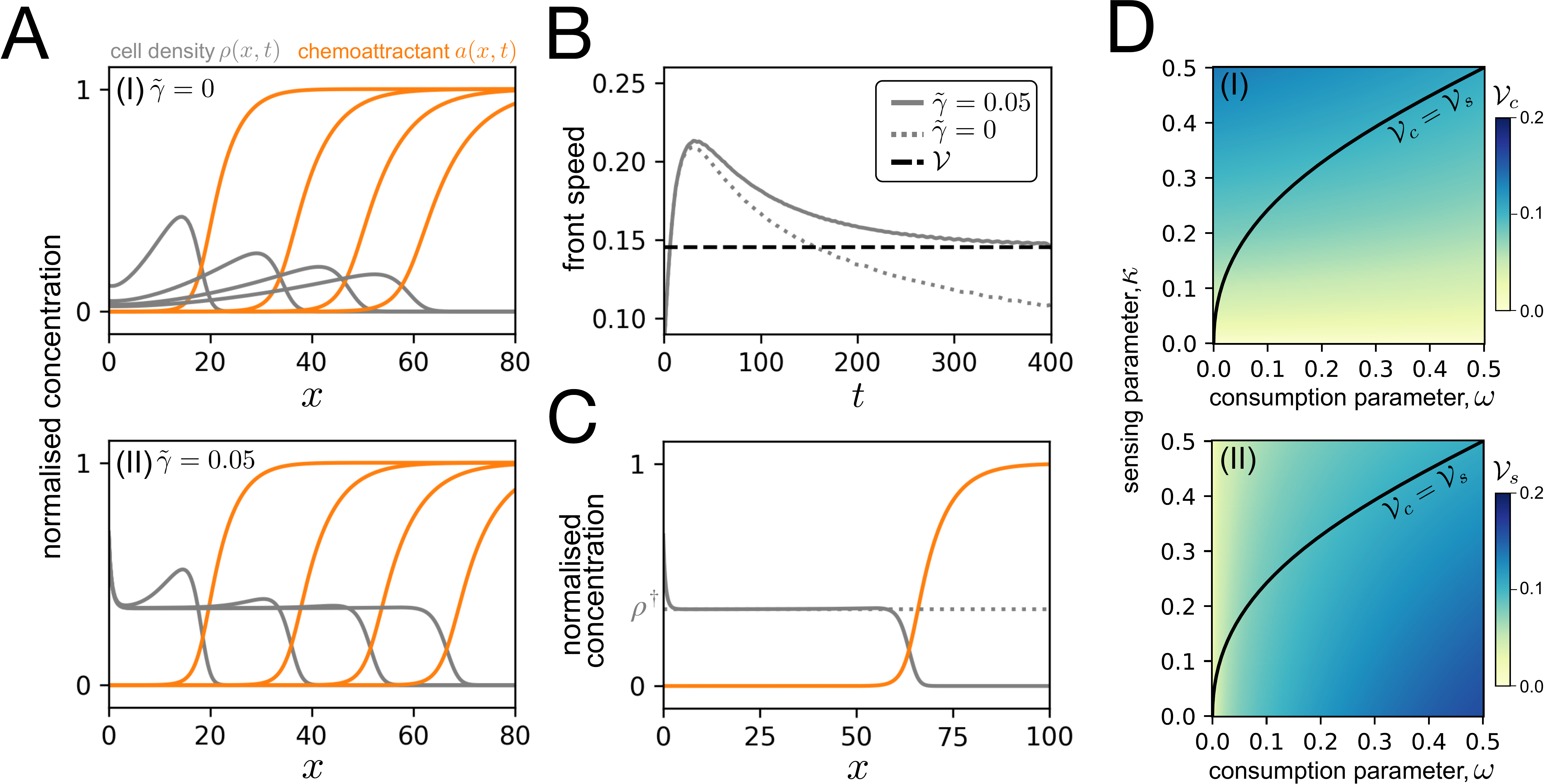}
    \caption{Travelling wave behaviour in the homogeneous migration model described by Equations~\eqref{eq:nondim_sys}--\eqref{eq:nondim_BCs_rho}. (A) Numerical travelling wave solutions shown at evenly spaced time points ($t=60$, $160$, $260$, and $360$), with cell density $\rho(x,t)$ shown in grey and chemoattractant concentration $a(x,t)$ shown in orange. In plot $(\sf{I})$, there is no cellular influx at the boundary ($\tilde{\gamma}=0$), whereas plot $(\sf{II})$ shows the case of sustained influx with $\tilde{\gamma}=0.05$. (B) Time evolution of the front speed for the simulations in (A). The black dashed line denotes the analytical prediction~\eqref{eq:homo_analysic_solns} for the travelling speed in the case $\tilde{\gamma}=0.05$. 
    (C) Comparison between the analytical predictions for the bulk cell density $\rho^\dagger$~\eqref{eq:homo_analysic_solns} and the long-term behaviour of the numerical simulations for $\tilde{\gamma}=0.05$. (D) Predicted travelling wave speeds for the migration of consumer and sensor cells in isolation~\eqref{eq:wave_speeds_homo}, with cell influx $\tilde\gamma = 0.05$. Subfigure $(\sf{I})$ shows the consumer population wave speed $\mathcal{V}_c$, while subfigure $(\sf{II})$ shows the sensor population wave speed $\mathcal{V}_s$. The black contour denotes the parameter combinations for which $\mathcal{V}_c = \mathcal{V}_s$. All other parameters are set to the values given in~\Cref{tab:params} for all subfigures.}
    \label{fig:FIG_homogeneous_system}
\end{figure}

Simulations of~\eqref{eq:nondim_sys_homo} reveal that two distinct migration regimes are possible.
The first occurs under a zero-flux boundary condition at the origin, \emph{i.e.}, $\tilde\gamma = 0$.
In this case, as the front advances, the profile gradually broadens into a near-flat distribution (Figure~\ref{fig:figure2A}({\sf I})), causing the propagation speed to decay to zero (Figure~\ref{fig:figure2B}).
In contrast, when a nonzero influx is imposed at the origin, \emph{i.e.}, $\tilde\gamma > 0$, the system develops a stable travelling wave: the cell density profile becomes monotonically decreasing (Figure~\ref{fig:figure2A}({\sf II})) and propagates toward $x \to  \infty$ at a constant speed (Figure~\ref{fig:figure2C}). 
The region behind the propagating wave, commonly referred to as \emph{the bulk}, retains a uniform cell density $\rho^{\dagger}$, while the chemotactic speed $\partial_x\log a$ also converges to a constant value $\lambda^\dagger$.
A small boundary layer forms near $x=0$, where the cell density rapidly decreases to converge to the homogeneous bulk concentration (\Cref{Figure5B}). This is an artefact of the Neumann boundary condition for the chemoattractant at $x=0$ and as such it is of no physical interest (further details can be found in Appendix~\ref{app:boundary_layer}).

\subsection{Derivation of the travelling wave speed}
\label{sec:TW homogeneous population}
Our aim is to characterise these travelling wave solutions analytically to understand how the different model parameters shape the invasion pattern of the population which requires studying non-linear eigenvalue problems for travelling wave solutions of \eqref{eq:nondim_sys_homo}. 
Considering the travelling wave scenario, \emph{i.e.}, $\tilde\gamma > 0$, we here revisit the derivation of the analytical expression of the travelling wave speed presented in~\cite{Ucar2025}.
By defining the travelling wave coordinate $z\in\mathbb{R}$, and looking for travelling wave solutions
\begin{equation}\label{eq:travelling_wave_coords_homo}
    \rho(x,t) = P(z), \quad a(x, t) = A(z) , \quad z = x - \mathcal{V} t,
\end{equation}
for Equations~\eqref{eq:nondim_sys_homo}, we obtain the system of ODEs 
\begin{subequations}\label{eq:homo_sys_TW}
    \begin{align}
        -\mathcal{V} P'(z) &= \tilde D_\rho P''(z) - \tilde\chi \left( P(z) \frac{A'(z)}{A(z)} \right)' , \label{eq:homo_sys_TW_a} \\
        -\mathcal{V} A'(z) &= A''(z) - P_c(z) A(z) \label{eq:homo_sys_TW_b} ,
    \end{align}
\end{subequations}
where the prime represents differentiation with respect to $z$.
Here we assume that $t$ is sufficiently large so that the behaviour near $x=0$ can be neglected.
Through numerical simulation, we observe that ahead of the wave front the cell population decays to zero, while in the bulk region, behind the front, the population density remains constant at some non-zero value $\rho^{\dagger}$ dependent on the value of certain system parameters. In contrast, the chemoattractant decays in the bulk region, while saturates at its maximum value $a=1$ at the front. We therefore couple~\eqref{eq:homo_sys_TW} to the following far-field conditions 
\begin{subequations}\label{eq:homo_sys_BCs_P}
\begin{equation}
    \lim_{z \to -\infty} (P,A) = (\rho^\dagger,0), \quad   \lim_{z \to +\infty} (P,A) = (0,1). 
\end{equation}
Importantly, in solving~\eqref{eq:homo_sys_TW}, we have to take care of the term $A'/A$ in~\eqref{eq:homo_sys_TW_a} which is singular as $A\to0$ in the bulk. Linearising Equation~\eqref{eq:homo_sys_TW_b} in the bulk region reveals that, at leading order, the chemoattractant $A(z)$ decays exponentially to zero in the bulk region $A\propto e^{\lambda^{\dagger}z}$, hence the singularity of $A'/A$ can be removed and
\begin{equation}\label{eq:homo_sys_BCs_A}
    \lim_{z\to-\infty} \frac{A'(z)}{A(z)} =\lambda^\dagger ,
\end{equation}
\end{subequations}
where the constant rate $\lambda^\dagger$, has to be positive to guarantee that the chemoattractant concentration decays to zero as $z\to-\infty$. Here, $\lambda^\dagger$ characterises the slope of the unstable manifold for the saddle point $(A,P,A',P')=(0,0,0,0)$.
Applying the asymptotic behaviour $A\propto e^{\lambda^{\dagger}z}$ as $z \to -\infty$ in \eqref{eq:homo_sys_TW_b} reveals that $\lambda^\dagger$ satisfies
\begin{equation}\label{eq:quadratic_lambda_homo}
    -\mathcal{V}\lambda^{\dagger} = (\lambda^{\dagger})^2 - \rho^\dagger.
\end{equation}

Integrating Equation~\eqref{eq:homo_sys_TW_a} over the domain $z \in (-\infty, +\infty)$, and applying boundary conditions in Equation~\eqref{eq:homo_sys_BCs_P}, yields
\begin{equation}
    \mathcal{V} \rho^\dagger = \tilde\chi \rho^\dagger \lambda^\dagger.\label{eq:def_lambda_dagger}
\end{equation}
Under the assumption $\rho^\dagger \neq 0$, which is satisfied for nonzero $\tilde\gamma$, this yields the expression
\begin{equation}\label{eq:vel_lam_relation_homo}
    \mathcal{V} = \tilde\chi \lambda^\dagger ,
\end{equation}
which reveals that the wave speed is directly proportional to the decay rate of the chemoattractant in the bulk.
This positive result for $\mathcal{V}$ is consistent with the interpretation of $a$ as a chemoattractant, since it confirms that the gradient drives migration in the direction of increasing $a$. As a result, the wave speed $\mathcal{V}$ is uniquely determined by the behaviour in the bulk, where the population stabilises and gradients in the cell density vanish. Using~\eqref{eq:quadratic_lambda_homo} and \eqref{eq:vel_lam_relation_homo}, we then remove dependence on the wave speed $\mathcal{V}$:
\begin{equation}\label{eq:homo_lambda_quadratic}
    (1 + \tilde\chi)(\lambda^{\dagger})^2 - \rho^\dagger = 0 .
\end{equation}
Therefore, we recover the same analytical expressions for the wave speed $\mathcal{V}$ and decay rate $\lambda^{\dagger}$ in terms of the system parameter $\tilde\chi$ and unknown bulk concentration $\rho^\dagger$ as in~\cite{Ucar2025}:
\begin{equation}
    \mathcal{V} =  \sqrt{ \frac{\rho^\dagger\tilde\chi^2}{1 + \tilde\chi} } \qquad \text{and} \qquad \lambda^{\dagger} = \sqrt{ \frac{\rho^\dagger}{1 + \tilde\chi} } .\label{eq:intermediate migration speed}
\end{equation}
Although the cell density in the bulk $\rho^{\dagger}$ might be a more accessible parameter from experimental observations, it is not an input of our model~\eqref{eq:nondim_sys_homo}. 
This suggests that there exists a family of travelling wave solutions for~\eqref{eq:nondim_sys_homo_a}--\eqref{eq:nondim_sys_homo_b} that are parametrised by $\rho^\dagger$. 
This raises the question of what mechanisms set the value of $\rho^{\dagger}$ in the numerical simulations of the full model~\eqref{eq:nondim_sys_homo}. 
This question was not addressed in~\cite{Ucar2025}, yet it is essential to understand the relationship between the migration dynamics, as described by travelling wave solutions, and the key model parameters: cell influx ($\tilde\gamma$), cell diffusion ($\tilde D_\rho$), and chemotactic sensitivity ($\tilde \chi$).

\subsubsection{Selection of the bulk cell concentration}\label{sec:homo_mass_conservation}
Informed by numerical simulations (Figure~\ref{fig:figure2A}({\sf II})), we see that the solution for $\rho$ settles towards a stationary profile near the inlet ($x=0$).
Hence, the quantity \begin{equation}
    \int_0^{x_0} \rho(x,t) \,\mathrm{d}x ,
\end{equation} quickly converges to a constant value in time, where $x_0$ is an arbitrary point in the bulk region behind the travelling wave front. 
Rewriting Equation~\eqref{eq:nondim_sys_homo_a} in the following conservative form
\begin{equation}\label{eq:rhoconservative}
    \diffp{\rho}{t} + \diffp{}{x} \left( \rho \diffp{}{x} \left[ - \tilde D_\rho \log{\rho} + \tilde\chi_\rho \log{a} \right] \right) = 0 ,
\end{equation}
and integrating it along the interval $x\in(0,x_0)$, we obtain
\begin{equation}\label{eq:cell_flux_homo}
    \frac{\mathrm{d}\int_0^{x_0}\rho \,\mathrm{d}x}{\mathrm{dt}}=\left[\rho \, \diffp{}{x} \left( - \tilde D_\rho \log{\rho} + \tilde\chi_\rho \log{a} \right)\right]^{x_0}_{0} \approx 0, \quad t\gg1.
\end{equation}
Physically, this implies that the rate at which cells enter the domain at $x=0$ is eventually equal to the rate at which cells are transported through the rear of the travelling wave at $x=x_0$. Hence, the total number of cells in the interval $[0,x_0]$ remains conserved.
Since the cell concentration is constant throughout the bulk region (\emph{i.e.}, $\partial_x \rho|_{x_0} \approx 0$) and the chemoattractant concentration decays exponentially with constant rate $\lambda^{\dagger}$, such that $\partial_x \log a|_{x_0} = \lambda^{\dagger}$, Equation~\eqref{eq:cell_flux_homo} yields the relation
\begin{equation}\label{eq:cell_flux_homo_2}
    \tilde\gamma = \tilde\chi_\rho \lambda^\dagger \rho^{\dagger}.
\end{equation}
Combining~\eqref{eq:intermediate migration speed} with~\eqref{eq:cell_flux_homo_2}, we can derive analytic expressions for the wave speed $ \mathcal{V}$, the chemoattractant decay rate $\lambda^{\dagger}$, and bulk concentration $\rho^\dagger$ solely in terms of model parameters:
\begin{equation}\label{eq:homo_analysic_solns}
    \mathcal{V} = \left( \frac{\tilde\gamma \tilde\chi^2}{1+\tilde\chi} \right) ^{1/3}, \qquad \lambda^\dagger = \left( \frac{\tilde\gamma}{\tilde\chi_\rho (1+\tilde\chi)} \right) ^{1/3}, \qquad \rho^\dagger = \left( \frac{\tilde\gamma^2 (1+\tilde\chi)}{\tilde\chi^2} \right) ^{1/3}.
\end{equation} 
In addition to yielding an explicit expression for the migration speed of travelling fronts~\eqref{eq:homo_analysic_solns}, our analysis reveals that the speed is fully determined by the behaviour in the bulk of the wave. 
This contrasts with other models of long-range migration, such as the Fisher--KPP equation, where the speed of travelling waves depends on the detailed structure of the leading edge~\cite{murray_biological_2002,simpson_fisherkpp-type_2024}. In the language of travelling wave theory, Fisher--KPP waves are being \emph{pulled}, whereas the waves in their system are being \emph{pushed}~\cite{Phillips2025}. 

A comparison of the analytic solutions in Equation~\eqref{eq:homo_analysic_solns} to the numerical solutions is presented in Figures~\ref{fig:figure2B}--\ref{fig:figure2C}. 
We find excellent agreement between the two, indicating that the assumptions underlying the analytic formulation are well justified and that the resulting solution accurately captures the behaviour of the system.

\subsection{Migration of consumer and sensor cells in isolation}
Having analysed the homogeneous population, we now examine the migration of the consumer and sensor cell populations in isolation. 
These cases serve as a baseline for comparison with the homogeneous model and provide insight into the intrinsic migration behaviour of each cell type before considering their interactions.
Setting $\rho_s(x,t) \equiv 0$ and then $\rho_c(x,t) \equiv 0$ in Equation~\eqref{eq:nondim_sys} and performing the same analysis as above yields the wave speeds for each population in the homogeneous state:
\begin{equation}\label{eq:wave_speeds_homo}
    \mathcal{V}_c = \left( \frac{(1-\omega)\tilde\gamma_c \, (\kappa \, \tilde\chi)^2}{1+\kappa \tilde\chi} \right) ^{1/3} \quad \text{and} \quad \mathcal{V}_s = \left( \frac{\omega\tilde\gamma_s \, ((1-\kappa) \, \tilde\chi)^2}{1+(1-\kappa) \tilde\chi} \right) ^{1/3} .
\end{equation}
Constraining the parameter space such that the wave speeds of the cell populations in isolation are equal, \emph{i.e.} $\mathcal{V}_c = \mathcal{V}_s$, requires
\begin{equation}\label{eq:kappa_omega_constraint}
    \omega = \frac{\tilde\gamma_c \kappa^2 (1 + (1-\kappa)\tilde\chi)}{\tilde\gamma_s(1-\kappa)^2 (1+\kappa \tilde\chi) + \tilde\gamma_c \kappa^2 (1+(1-\kappa)\tilde\chi)} .
\end{equation}
Figure~\ref{fig:figure2D} depicts the wave speeds of the two populations in isolation when varying heterogeneity parameters $\omega$ and $\kappa$, with the trajectory $\mathcal{V}_c = \mathcal{V}_s$ illustrated in black.
The boundary is monotonically increasing, indicating a continuous trade-off between sensing and consumption. 
Populations with a reduced capacity to sense the chemotactic gradient can maintain the same migration speed by increasing their consumption rate, while populations with weaker consumption can compensate through enhanced sensing. 
Thus, neither mechanism alone determines migration performance; instead, it is the balance between gradient generation and gradient sensing that governs the travelling wave speed.

Overall, the homogeneous analysis provides a clear baseline description of self-generated chemotactic invasion in the absence of heterogeneity. 
By deriving explicit relationships between the travelling wave properties and the underlying model parameters, we obtain direct insight into the mechanisms controlling wave propagation and chemoattractant depletion. 
These results not only extend the analytical framework of \cite{Ucar2025}, but also establish the foundation for understanding how the introduction of a second interacting population modifies the invasion dynamics in the heterogeneous system.

\begin{figure}
    \centering
    \begin{subfigure}{0.0\textwidth}
        \captionlistentry{}
        \label{fig:figure3A}
    \end{subfigure}
        \begin{subfigure}{0.0\textwidth}
        \captionlistentry{}
        \label{fig:figure3B}
    \end{subfigure}
        \begin{subfigure}{0.0\textwidth}
        \captionlistentry{}
        \label{fig:figure3C}
    \end{subfigure}
        \begin{subfigure}{0.0\textwidth}
        \captionlistentry{}
        \label{fig:figure3D}
    \end{subfigure}
        \begin{subfigure}{0.0\textwidth}
        \captionlistentry{}
        \label{fig:figure3E}
    \end{subfigure}
    \includegraphics[width=\linewidth]{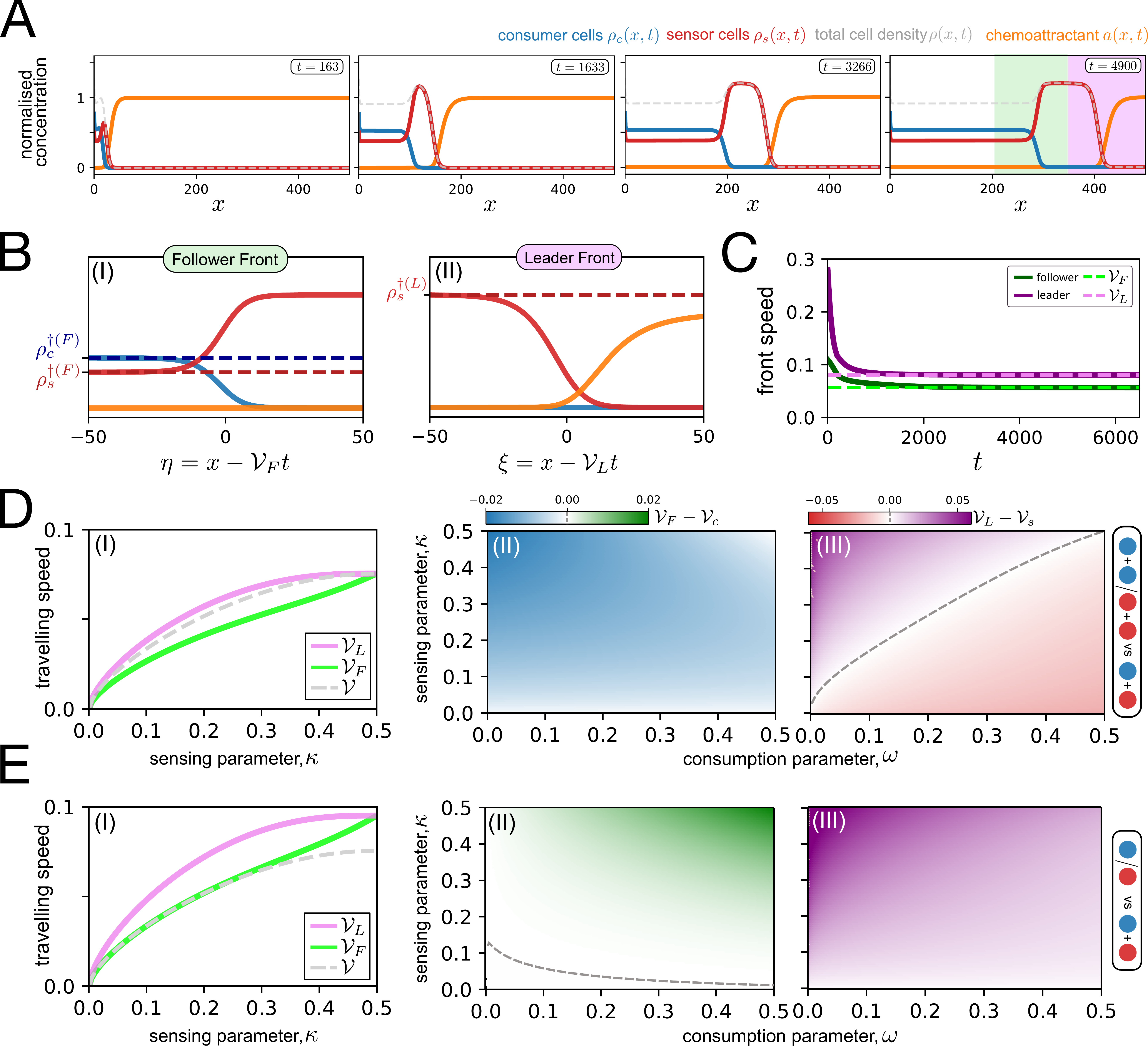}
    \caption{(A) Time snapshots of collective migration of mixed sensor-consumer cells. The profiles are obtained by numerically solving Equations~\eqref{eq:nondim_sys}--\eqref{eq:nondim_BCs_rho}. Time increases from left to right. In the last panel, the green and pink shaded areas indicate the follower and leader fronts, respectively. (B) Structure of the asymptotic propagating terrace solution. Continuous curves are as in the last panel of~\Cref{fig:figure3A}. Horizontal dotted lines correspond to the analytically estimated values of the bulk concentrations ($\rho_c^{\dagger(F)}$, $\rho_s^{\dagger(F)}$, and $\rho_s^{\dagger(L)}$) implicitly defined by~\eqref{eq:sim_eqns_two_waves}. (C) Comparison between the speeds at which the two fronts in~\Cref{fig:figure3A} propagate and the analytical estimates of the travelling wave speeds associated with propagating terrace solutions of~\eqref{eq:nondim_sys}. (D) and (E) present the change in invasion speed arising from co-migration in the heterogeneous system, compared to isolated migration. (D) Phenotypic specialisation within a population: the total number of cells is equal to the homogeneous benchmark, \emph{i.e.} $\tilde\gamma_c = \tilde\gamma_s = \frac{1}{2}\tilde\gamma$. (E) Co-migration of two distinct populations: the total number of cells is double that of the homogeneous benchmark, \emph{i.e.} $\tilde\gamma_c = \tilde\gamma_s = \tilde\gamma$. ({\sf I}) Comparison of travelling wave speeds along the contour $\mathcal{V}_c = \mathcal{V}_s$, when the cell influx of the homogeneous case is $\tilde\gamma=0.05$. ({\sf II}) Shows the change in front propagation speed for the consumer cells ($\mathcal{V}_F - \mathcal{V}_c$) and ({\sf III}) shows the change for sensor cells ($\mathcal{V}_L - \mathcal{V}_s$), as heterogeneity parameters $\omega$ and $\kappa$ are varied. In all subfigures, the remaining parameters are set to the values given in~\Cref{tab:params}.}
    \label{fig:FIG_heterogeneous_system}
\end{figure}

\section{Coupled migration of sensor-consumer cell mixtures}
\label{sec:heterogeneous chemotaxis}
Having characterised the migration of the homogeneous populations, we now return to the full heterogeneous model defined by Equations~\eqref{eq:nondim_sys}--\eqref{eq:nondim_BCs_rho} and investigate how heterogeneity affects chemotactic migration.
Numerical solutions display a characteristic two–front structure, which we denote as leading and following fronts (Figures~\ref{fig:figure3A}--\ref{fig:figure3C}). These simulations are performed in the regime where isolated sensor and consumer populations propagate at the same travelling wave speed (Equation~\eqref{eq:kappa_omega_constraint}). This suggests that separation into two travelling fronts does not arise because one cell type is intrinsically faster than the other. Rather, the emergence of two distinct fronts is a consequence of coupling the two populations through the shared chemoattractant.

The following front is characterised by the depletion of the consumer population, while the sensor population rises and approaches a new bulk density $\rho_s^{\dagger(L)}$ ahead of the front. This front moves at speed $\mathcal{V}_F$ into the region only occupied by sensor cells.
The faster leading front propagates ahead of this region where the sensor population rapidly decays to zero at a speed $\mathcal{V}_L$ (Figure~\ref{fig:figure3C}). The two fronts partition the domain into three regions in which the cell densities are constant (Figure~\ref{fig:figure3B}). Behind the following front, both consumer and sensor populations are present with densities $\rho_c^{\dagger(F)}$ and $\rho_s^{\dagger(F)}$, respectively. Between the two fronts, the consumer cells vanish, while the sensor cells reach their maximal bulk density $\rho_s^{\dagger(L)}$. Finally, ahead of the leading front,  both consumer and sensor cell densities vanish. Together, these fronts produce a spatially ordered pattern with distinct bulk regions separated by well-defined travelling interfaces that eventually move infinitely far apart. 
Such solutions are also known as \emph{propagating terraces}---a term commonly used in the reaction-diffusion literature to describe general frontal behaviours in multi-stable systems \cite{Ducrot2012ExistenceAC}.

Informed by the results of our numerical simulations, we now seek to characterise the invasion pattern observed in simulations of heterogeneous consumer/sensor populations across parameter regimes. We do so by investigating propagating terrace solutions of Equations~\eqref{eq:nondim_sys}.
As with the homogeneous system, we focus on characterising analytically the travelling speed of the two fronts in terms of model parameters. 
This approach helps explain the migration patterns observed in simulations, investigate their robustness across parameter regimes, and connect them directly to the different physical mechanisms included in the model.

\subsection{Derivation of propagating terrace speeds}
Our aim is to characterise propagating terrace solutions of Equations~\eqref{eq:nondim_sys} analytically. At long times, the boundary layer near $x=0$ remains localised while the two fronts separate linearly in time (Figure~\ref{fig:figure3A}). We therefore neglect the boundary layer and analyse the asymptotic regime in which the fronts are effectively infinitely separated. Hence, we study the two fronts independently by introducing the two independent travelling wave coordinates (Figure~\ref{fig:figure3B})
\begin{equation}\label{eqs:travelling wave coordinates}
    \eta = x - \mathcal{V}_Ft,\quad \xi = x - \mathcal{V}_Lt, \quad \xi,\eta\in\mathbb{R},
\end{equation}
then matching the two solutions in the intermediate region ($\xi\to-\infty$ and $\eta\to \infty$). This is trivial for the cell distributions $\rho_s$ and $\rho_c$---which converge to constant values. 
In contrast, the limiting behaviour of the chemoattractant profile in the intermediate region is different in the two travelling wave reference frames. 
Specifically, it assumes a non-stationary profile in the follower reference frame. Accounting for this behaviour allows us to derive a coupled system of algebraic equations determining the wave speeds and bulk compositions.

\subsubsection{Characterisation of the leading front}\label{sec:leading_front}
We begin by focusing on the structure of the solution near the leading front by seeking travelling wave solutions of the form
\begin{equation}\label{eq:travelling_wave_coords_2}
    \rho_c(x,t) = P_c(\xi), \quad \rho_s(x,t) = P_s(\xi), \quad a(x, t) = A(\xi), \quad \xi = x - \mathcal{V}_Lt ,
\end{equation}
where $\mathcal{V}_L$ indicates the wave speed of the leading front.
Substituting \eqref{eq:travelling_wave_coords_2} into \eqref{eq:nondim_sys} yields the following system of coupled boundary value problems:
\begin{subequations}\label{eq:front_2_ODEs}
    \begin{align}
        -\mathcal{V}_L P_s'(\xi) &= \tilde D_s P_s''(\xi) - (1-\kappa) \tilde\chi \left( P_s(\xi) \frac{A'(\xi)}{A(\xi)} \right)' \label{eq:front_2_ODEs_a} ,\\
         -\mathcal{V}_L P_c'(\xi) &= \tilde D_c P_c''(\xi) - \tilde\chi_c \left( P_c(\xi) \frac{A'(\xi)}{A(\xi)} \right)' \label{eq:front_2_ODEs_b} ,\\
        -\mathcal{V}_L A'(\xi) &= A''(\xi) - ((1-\omega)P_c(\xi)+\omega P_s(\xi)) A(\xi) , \label{eq:front_2_ODEs_c}
    \end{align}
\end{subequations}
where the primes represent differentiation with respect to the travelling wave coordinate $\xi$.
As illustrated in Figure~\ref{fig:figure3B}({\sf II}), we look for solutions that connect the bulk region in which the cell density is non-zero to the front region in which the cell population decays to zero $(0,0,1)$ via imposing the far-field conditions
\begin{equation}\label{eq:front_2_BSc_cells}
    \lim_{\xi \to -\infty} (P_c,P_s,A) = (\rho_c^{\dagger(L)},\rho_s^{\dagger(L)},0) \quad   \lim_{\xi \to +\infty} (P_c,P_s,A) = (0,0,1).
\end{equation}
Integrating~\eqref{eq:front_2_ODEs}, we find that, provided $\kappa\neq0.5$, one of the two cell types has to vanish in the bulk, \emph{i.e.}, $\rho_c^{\dagger(L)}\rho_s^{\dagger(L)}$. Details are given in~\Cref{app:tw_analysis} of the Appendix. Informed by the dynamical simulations (\Cref{fig:figure3A}), we focus on the scenario in which $\rho_s^{\dagger(L)}\neq 0$ and $\rho_c^{\dagger(L)}=0$. Given that the consumer population vanishes both at the front and the back of the wave, $P_c(\xi) \equiv 0$ is a solution to~\eqref{eq:front_2_ODEs_b}. 

Substituting $P_c(\xi) \equiv 0$ into \eqref{eq:front_2_ODEs}, we recover the same travelling wave problem as the one discussed in Section~\ref{sec:TW homogeneous population}---up to redefinition of model parameters. Therefore, we can directly apply Equation~\eqref{eq:homo_lambda_quadratic} to obtain:
\begin{equation}\label{eq:speed_leading}
    \mathcal{V}_L = (1-\kappa) \tilde\chi \lambda_L,
\end{equation}
and
\begin{equation}\label{eq:sim_eq_1}
    \lambda_L^2(1 - (1-\kappa) \tilde\chi ) - \omega \rho_s^{\dagger(L)} = 0,
\end{equation}
where $\lambda_L>0$ is the rate at which the solution approaches the steady state $(0,\rho_s^{\dagger(L)},0)$ as $\xi\to-\infty$.
Here, we can not apply the same mass conservation argument as in Section~\ref{sec:homo_mass_conservation} to estimate $\rho_s^{\dagger(L)}$ as profile behind the bulk is no longer constant owing to the presence of the slower following front.

We conclude that, like before, the speed of the leading front is proportional to its own chemotactic ability, however, as we shall see, the bulk concentration $\rho_s^{\dagger(L)}$ is influenced by the spatial structure of the solution near the following front, reflecting non-local effects beyond the direct contribution from the influx at the origin. 
In other words, the presence and shape of the following front modulates the bulk concentration of the leading population, reflecting the interdependence of the two fronts in the heterogeneous system.

\subsubsection{Characterisation of the following front}
Having characterised the behaviour of the leading front, we now focus on the following front. 
The crucial novelty in the analysis of the following front is the non-steady nature of the chemoattractant profile owing to the presence of the faster-moving leading front. 
As a result, the profile of $a$ at the following front depends not only on the local dynamics; rather, it is coupled to the behaviour of the leading front further ahead.

In the travelling wave coordinate of the following front $\eta$~\eqref{eqs:travelling wave coordinates}, the behaviour of the chemoattractant profile in the intermediate region $\eta\to \infty$ becomes
\begin{equation}\label{eq:matching_chemoattractant}
   A(\eta,t)\propto e^{\lambda_L \xi} = e^{\lambda_L(\eta + (\mathcal{V}_F-\mathcal{V}_L)t)} = e^{\lambda_L \eta}e^{\lambda_L(\mathcal{V}_F-\mathcal{V}_L)t}, \quad \xi\to-\infty ,
\end{equation}
where the temporal decay of the chemoattractant is dependent on the difference in wave speeds of the two fronts. The matching condition~\eqref{eq:matching_chemoattractant} suggests seeking a separable form for $A$:
\begin{equation}
    A(\eta, t) = e^{\lambda_L(\mathcal{V}_F-\mathcal{V}_L)t} \widetilde A(\eta) ,
\end{equation}
with $\widetilde A(\eta)$ being a time-invariant profile that grows exponentially $\widetilde A(\eta)\sim e^{\lambda_L\eta}$ as $\eta\to\infty$.

We therefore seek travelling wave solutions of the form
\begin{equation}\label{eq:travelling_wave_coords_1}
A(\eta,t) = e^{\lambda_L(\mathcal{V}_F - \mathcal{V}_L)t} \widetilde A(\eta), \qquad P_c=P_c(\eta), \qquad P_s=P_s(\eta), \qquad \eta = x - \mathcal{V}_F t.
\end{equation}
Substituting~\eqref{eq:travelling_wave_coords_1} into~\eqref{eq:nondim_sys}, we obtain the following system of coupled ODEs
\begin{subequations}\label{eq:front_1_ODEs}
    \begin{align}
        -\mathcal{V}_F P_c'(\eta) &= \tilde D_c P_c''(\eta) - \kappa \tilde\chi \left( P_c(\eta) \frac{\widetilde A'(\eta)}{\widetilde A(\eta)} \right)' \label{eq:front_1_ODEs_a} ,\\
        -\mathcal{V}_F P_s'(\eta) &= \tilde D_s P_s''(\eta) - (1-\kappa) \tilde\chi \left( P_s(\eta) \frac{\widetilde A'(\eta)}{\widetilde A(\eta)} \right)' \label{eq:front_1_ODEs_b}, \\
        \lambda_L (\mathcal{V}_F - \mathcal{V}_L) \tilde A(\eta) -\mathcal{V}_F \tilde A'(\eta) &= \tilde A''(\eta) - \Big[ \omega P_s(\eta) + (1-\omega) P_c(\eta) \Big] \tilde A(\eta) \label{eq:front_1_ODEs_c} ,
     \end{align}
\end{subequations} 
where the primes represent differentiation with respect to $\eta$.
From numerical simulations, depicted in Figure~\ref{fig:figure3B}({\sf I}), we seek for solutions that connect the bulk region of the following front, where sensor and consumer cells are mixed, to the bulk region of the leading front~\eqref{eq:front_2_BSc_cells}, where the consumer cell population decays to zero while the sensor cells and $\widetilde A$ approach a constant non-zero value
\begin{equation}\label{eq:front_1_BCs_cells}
    \lim_{\eta \to -\infty} (P_c, P_s,\widetilde{A})= (\rho_c^{\dagger(F)},\rho_s^{\dagger(F)},0) \quad \lim_{\eta \to \infty} (P_c, P_s,\widetilde{A})= (0,\rho_s^{\dagger(L)},e^{\lambda_L\eta}). 
\end{equation}
These boundary conditions prescribe the asymptotic (bulk) values of the unknown travelling wave profiles far behind $\rho_{c,s}^{\dagger(F)}$ and far ahead of the front $\rho_{s}^{\dagger(L)}$, which are all as-of-yet unknown. 

Again, we can integrate~\eqref{eq:front_1_ODEs_a}--\eqref{eq:front_1_ODEs_b} exactly; by imposing the far-field condition~\eqref{eq:front_1_BCs_cells} at $\eta\to \infty$, we obtain:
\begin{subequations}\label{eq:front_1_ODEs_one_integration}
    \begin{align}
        -\mathcal{V}_F P_c(\eta) &= \tilde D_c P_c'(\eta) - \kappa \tilde\chi P_c(\eta) \frac{\widetilde A'(\eta)}{\widetilde A(\eta)}  \label{eq:front_1_ODEs_a_integration} ,\\
        -\mathcal{V}_F P_s(\eta) &= \tilde D_s P_s'(\eta) - (1-\kappa) \tilde\chi P_s(\eta) \frac{\widetilde A'(\eta)}{\widetilde A(\eta)} +(1-\kappa) \tilde\chi \rho^{\dagger(L)}_s(\eta)\lambda_L\label{eq:front_1_ODEs_b_integration}, \\
        \lambda_L (\mathcal{V}_F - \mathcal{V}_L) \tilde A(\eta) -\mathcal{V}_F \tilde A'(\eta) &= \tilde A''(\eta) - \Big[ \omega P_s(\eta) + (1-\omega) P_c(\eta) \Big] \tilde A(\eta) \label{eq:front_1_ODEs_c_integration} ,
     \end{align}
\end{subequations} 
As in~\Cref{sec:TW homogeneous population}, the travelling wave speed $\mathcal{V}_F$ can be computed by linearising~\eqref{eq:front_1_ODEs_one_integration} in the bulk ($\eta\to -\infty$) with an exponential ansatz:
\begin{equation}\label{eq:front1-far-field}
(P_c,P_s,\widetilde A)= (\rho_c^{\dagger(F)},\rho_s^{\dagger(F)},0) + \delta\vec{u}\, e^{\lambda_F\eta},
\end{equation}
where $\lambda_F>0$ represents the decay rate of the chemoattractant behind the following front and the vector $\delta\vec{u}\in\mathbb{R}^3$ is small, \emph{i.e.}, $|\delta\vec{u}|\ll1$.
Substituting~\eqref{eq:front1-far-field} into Equations~\eqref{eq:front_1_ODEs_a_integration}--\eqref{eq:front_1_ODEs_b_integration} we obtain
\begin{equation}\label{eq:speed_following}
    \mathcal{V}_F = \kappa \tilde\chi \lambda_F ,
\end{equation}
from \eqref{eq:front_1_ODEs_a}, and
\begin{equation}\label{eq:sensor_front_1_equation}
    \mathcal{V}_F \Big( \rho_s^{\dagger(L)} - \rho_s^{\dagger(F)} \Big) =  (1-\kappa) \tilde\chi \Big( \rho_s^{\dagger(L)} \lambda_L - \rho_s^{\dagger(F)} \lambda_F \Big) ,
\end{equation}
from \eqref{eq:front_1_ODEs_b}, which becomes
\begin{equation}\label{eq:sim_eq_2}
    \kappa \lambda_F \Big( \rho_s^{\dagger(L)} - \rho_s^{\dagger(F)} \Big) =  (1-\kappa) \Big( \rho_s^{\dagger(L)} \lambda_L - \rho_s^{\dagger(F)} \lambda_F \Big) ,
\end{equation}
after substitution of Equation~\eqref{eq:speed_following} to remove dependence on $\mathcal{V}_F$.
Linearising Equation~\eqref{eq:front_1_ODEs_c} we then find an expression for $\lambda_F$ 
\begin{equation}\label{eq:lambda_F_quadratic}
    \lambda_L(\mathcal{V}_F - \mathcal{V}_L) = \lambda_F^2 + \mathcal{V}_F \lambda_F - \Big( \omega \rho_s^{\dagger(F)} + (1-\omega) \rho_c^{\dagger(F)} \Big) .
\end{equation}
Using the identities~\eqref{eq:speed_leading} and \eqref{eq:speed_following} to remove the dependence of~\eqref{eq:lambda_F_quadratic} on the wave speeds, we find
\begin{equation}\label{eq:sim_eq_3}
    \lambda_F^2 (1+\kappa \tilde\chi) + \tilde\chi \lambda_L \Big((1-\kappa)\lambda_L - \kappa\lambda_F \Big) - \Big( \omega \rho_s^{\dagger(F)} + (1-\omega) \rho_c^{\dagger(F)} \Big) = 0 .
\end{equation}
To summarise, we have derived five Equations~\eqref{eq:speed_leading}, \eqref{eq:sim_eq_1}, \eqref{eq:speed_following}, \eqref{eq:sim_eq_2}, and \eqref{eq:sim_eq_3} relating the (as-of-yet unknown) quantities $\mathcal{V}_F$, $\mathcal{V}_L$, $\rho_c^{\dagger(F)}$, $\rho_s^{\dagger(F)}$, $\rho_s^{\dagger(L)}$, $\lambda_F$, and $\lambda_L$ to system parameters. 
The wave speed Equations~\eqref{eq:speed_leading} and \eqref{eq:speed_following} are independent of the remaining system and determine $\mathcal{V}_L$ and $\mathcal{V}_F$. 
The problem therefore reduces to five unknowns and three equations, hence only two of the bulk densities can be arbitrarily set, while the remaining is prescribed by the coupling of the two fronts.
Again, we observe a direct coupling between the two propagating fronts. 
The leading front influences the trailing front by shaping the chemoattractant gradient, while the trailing front, in turn, affects the leading front by influencing the bulk concentration $\rho_s^{\dagger(L)}$.

\subsubsection{Selection of the bulk cell concentrations}
The unknown quantities to be determined are the bulk concentrations, ($\rho_c^{\dagger(F)}$, $\rho_s^{\dagger(F)}$, and $\rho_s^{\dagger(L)}$), together with the decay rates, ($\lambda_F$ and $\lambda_L$), which together determine the wave speeds of the following and leading fronts through \eqref{eq:speed_following} and \eqref{eq:speed_leading}, respectively.
At this stage, we have three equations for five unknowns, given by Equations~\eqref{eq:sim_eq_1}, \eqref{eq:sim_eq_2}, and \eqref{eq:sim_eq_3}. 
Two additional equations can be obtained by applying mass conservation arguments, balancing the fluxes of cells at the rear of the following front with the one imposed at the boundary $x=0$. 
Importantly, this argument is only valid for the following front and not the leading front, where the bulk cell concentrations are set by matching the behaviour of the two fronts.

As in Section~\ref{sec:homo_mass_conservation}, we write Equations~\eqref{eq:nondim_sys_a}--\eqref{eq:nondim_sys_b} in conservative form to yield expressions for the flux of each cell population:
\begin{subequations}
    \begin{align}
        F_c(x, t) &= \rho_c \, \diffp{}{x} \left[ - \tilde D_c \log{\rho_c} + \kappa \tilde\chi \log{a} \right] = -\tilde D_c \diffp{\rho_c}{x} + \kappa \tilde\chi \rho_c \diffp{}{x} \Big[ \log{a} \Big] , \\
        F_s(x, t) &= \rho_s \, \diffp{}{x} \left[ - \tilde D_s \log{\rho_s} + (1-\kappa)\tilde\chi \log{a} \right] = -\tilde D_s \diffp{\rho_s}{x} + (1-\kappa) \tilde\chi \rho_s \diffp{}{x} \Big[ \log{a} \Big] ,
    \end{align}
\end{subequations}
and evaluate these expressions in the bulk region behind the following front,
\begin{subequations}\label{eq:flux_balance_hetero}
    \begin{align}
        \tilde \gamma_c &= \kappa \tilde\chi \lambda_F \rho_c^{\dagger(F)} \label{eq:flux_balance_hetero_a}\\
        \tilde \gamma_s &= (1-\kappa) \tilde\chi \lambda_F \rho_s^{\dagger(F)} \label{eq:flux_balance_hetero_b}.
    \end{align}
\end{subequations}

Equations~\eqref{eq:sim_eq_1}, \eqref{eq:sim_eq_2}, \eqref{eq:sim_eq_3}, \eqref{eq:flux_balance_hetero_a}, and \eqref{eq:flux_balance_hetero_b} constitute a closed system of five independent algebraic equations for the five unknown variables $\rho_c^{\dagger(F)}$, $\rho_s^{\dagger(F)}$, $\rho_s^{\dagger(L)}$, $\lambda_F$, and $\lambda_L$, that they uniquely define in terms of system parameters:
\begin{subequations}\label{eq:sim_eqns_two_waves}
    \begin{align}
        0 &= \lambda_L^2(1+(1-\kappa) \tilde\chi) - \omega \rho_s^{\dagger(L)} , \label{eq:sim_eqns_two_waves_a}\\
        0 &= \lambda_F^2 (1+\kappa \tilde\chi) + \tilde\chi \lambda_L ((1-\kappa)\lambda_L - \kappa\lambda_F ) - ( \omega \rho_s^{\dagger(F)} + (1-\omega) \rho_c^{\dagger(F)} ) , \label{eq:sim_eqns_two_waves_b}\\
        0 &= (2\kappa - 1) \rho_s^{\dagger(F)} \lambda_F + ((1-\kappa) \lambda_L - \kappa \lambda_F) \rho_s^{\dagger(L)} ,\\
        0 &= \kappa \tilde\chi \lambda_F \rho_c^{\dagger(F)} - \tilde\gamma_c ,\\
        0 &= (1-\kappa) \tilde\chi \lambda_F \rho_s^{\dagger(F)} - \tilde\gamma_s .
    \end{align}
\end{subequations}
Since we have quadratic expressions for decay rates $\lambda_F$ and $\lambda_L$, we require the roots of the corresponding quadratic equations~\eqref{eq:sim_eqns_two_waves_a}--\eqref{eq:sim_eqns_two_waves_b} to satisfy $\lambda_F, \lambda_L > 0$.
Based on the propagating terrace analysis, we can also conclude that the leading front is a pushed wave~\cite{Phillips2025}---as for the homogeneous case (\Cref{sec:homo_mass_conservation}). 
In contrast, the  following front does not fall within the standard pulled-pushed wave dichotomy since its speed depends both on the behaviour of the solution ahead and behind the travelling front.

Although we cannot write explicit solutions to the non-linear system in \eqref{eq:sim_eqns_two_waves}, we can approximate its roots numerically using the \texttt{fsolve} function from the \texttt{scipy.optimize} python package. 
As shown in Figure~\ref{fig:figure3B} the analytic prediction for the bulk concentration values and wave speeds are in excellent agreement with the long term behaviour obtained from numerical simulations of \eqref{eq:nondim_sys}, confirming that our analytic approach reliably captures the system's long-time behaviour. Hence,~\eqref{eq:sim_eqns_two_waves} allows us to efficiently study how the coupled migration of sensor-consumer cell population changes with model parameters without having to solve the full time dependent problem~\eqref{eq:nondim_sys}.

\subsection{Heterogeneity modifies invasion speeds through coupled chemotactic gradients}
We use the analytical predictions for the propagating terrace wave speeds to quantify how sensor/consumer heterogeneity affects long-range migration. 
As a first comparison, we match the total cell influx in the heterogeneous and homogeneous systems by setting $\tilde{\gamma}_c+\tilde{\gamma}_s=\tilde{\gamma}$. 
Figure~\ref{fig:figure3D}({\sf I}) compares the homogeneous travelling wave speed $\mathcal{V}$ with the speed of the leading ($\mathcal{V}_L$) and following ($\mathcal{V}_F$) fronts in the propagating terrace, restricting to the parameter combinations satisfying~\eqref{eq:kappa_omega_constraint}, for which sensor and consumer homogeneous populations have the same migration speed $\mathcal{V}$. 
We find that the speeds of the leading ($\mathcal{V}_L$) and following ($\mathcal{V}_F$) fronts consistently deviate from $\mathcal{V}$, especially for intermediate values of $\kappa$. 
While the leading front moves faster than the homogeneous population, the following front that consists of a mixture of sensor/consumer cells propagates slower than the homogeneous benchmark. 
This suggests that coupling of sensing and consumer cells changes the invasion dynamics, even when homogeneous sensor and consumer cell populations migrate at the same speeds. 
In particular, our results show that coupling provides an advantage for sensor cells while it penalises consumer cells that are confined to the following front. 
Extending the comparison across the full $(\kappa,\omega)$-parameter space (Figures~\ref{fig:figure3D}({\sf II})-({\sf III})) reveals that consumer cells always migrate more slowly in heterogeneous populations compared to homogenous. 
In contrast, the benefit to sensor cells depends on their intrinsic migration ability. 
When sensor cells migrate poorly as a collective, coupling to the consumer population substantially enhances their invasion speed. 
As their intrinsic migration ability increases, however, this advantage progressively diminishes. 
Beyond a critical threshold (gray dashed curve in  Figure~\ref{fig:figure3D}({\sf III})), sensor cells migrate faster in homogeneous populations than in heterogeneous mixtures.

As a complementary comparison, we now assign each heterogeneous subpopulation the same boundary influx as its corresponding homogeneous population, so that $\tilde{\gamma}=\tilde{\gamma}_c=\tilde{\gamma}_s$. 
We find that, generally, both the following and leading travelling fronts forming in the mixing scenario exceed the homogeneous benchmark across parameter values satisfying~\eqref{eq:kappa_omega_constraint}. 
This supports the main interpretation that co-migration through a jointly depleted chemoattractant can enhance the migration speed of the collective. 
Yet, the increased migratory ability is not simply a trivial consequence of adding densities. Its consequences depends on how sensing strength, consumption, and boundary influx combine to set the coupled bulk states and chemoattractant gradients. 
Across the full $(\kappa,\omega)$-parameter space, sensor cells migrate faster in a heterogeneous population rather than in isolation (Figure~\ref{fig:figure3E} ({\sf II})). 
For the consumer cells the same holds across most of the parameter space, except for low values of $\kappa$. 
Hence, unless the sensing ability of the consumer cells is significantly lower than that of the sensor cells, consumer cells will benefit from migrating with sensor cells rather than in isolation. 
Yet, based on the result in Figure~\ref{fig:figure3D} ({\sf II})), they would still benefit the most from interacting with purely other consumer cells as that would yield overall steeper gradients.

\begin{figure}[htp]
    \centering
    \includegraphics[width=\linewidth]{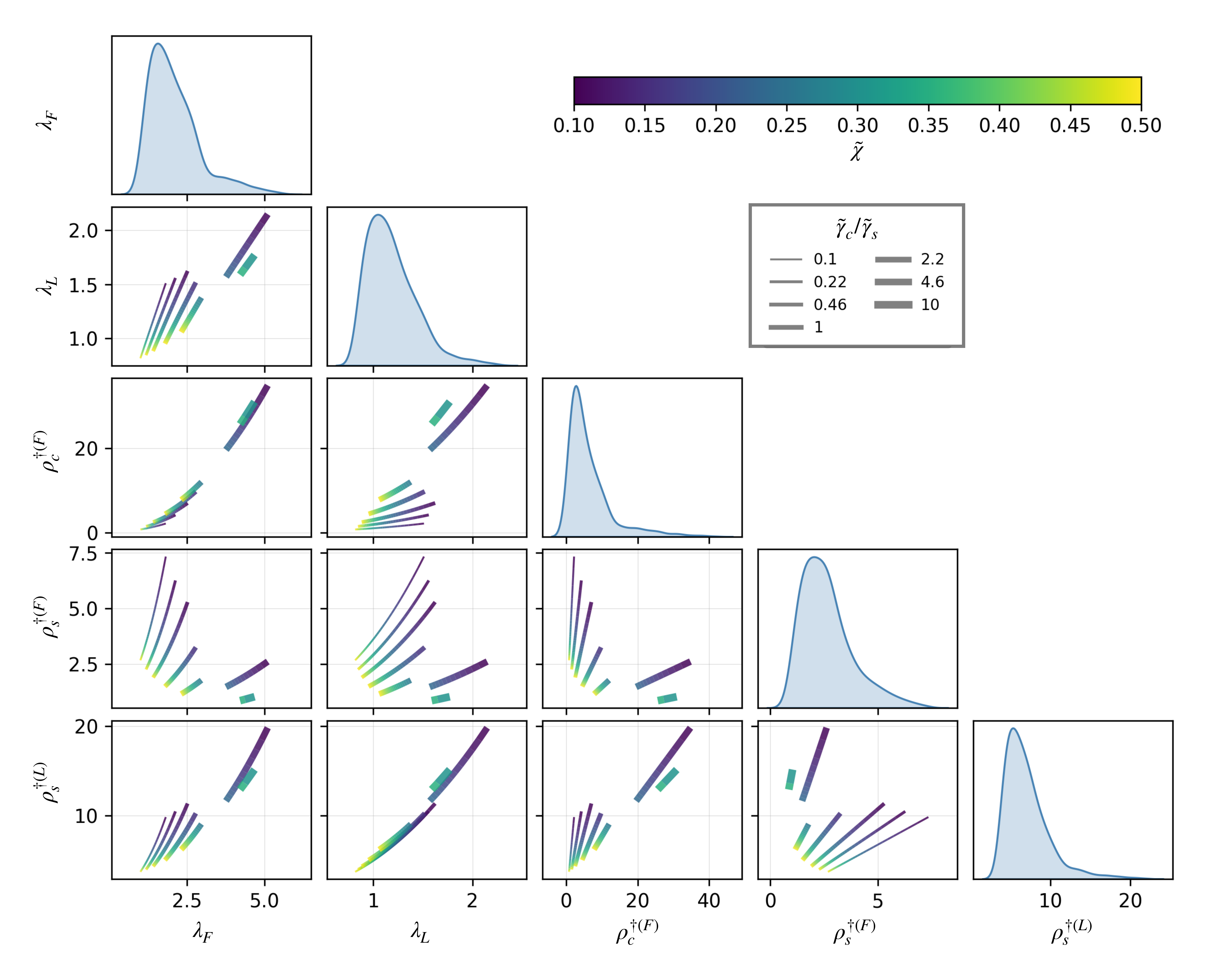}
    \caption{Pairwise projections of the converged solutions to Equation~\eqref{eq:sim_eqns_two_waves} for varying interaction strength $\tilde{\chi}$ and flux ratio $\tilde{\gamma}_c/\tilde{\gamma}_s$. The diagonal plots show the marginal distributions of each variable, while the off-diagonal plots show pairwise correlations between the chemoattractant decay rates $(\lambda_F,\lambda_L)$ and bulk concentrations $(\rho_c^{\dagger(F)},\rho_s^{\dagger(F)},\rho_s^{\dagger(L)})$. Rather than forming diffuse clouds, the solutions lie along narrow curved trajectories, indicating that the admissible states are strongly constrained by the coupled nonlinear equations. The emergence of these low-dimensional nonlinear manifolds demonstrates the strong interdependence between all dynamical variables. All other parameters are set to the values given in Table~\ref{tab:params}.}
    \label{fig:FIG_heterogeneous_param_dependence}
\end{figure}

To further highlight the interdependence of the two wave fronts, Figure~\ref{fig:FIG_heterogeneous_param_dependence} demonstrates the dependence of the wave speeds and bulk concentrations on the governing parameters $\tilde{\chi}$ and $\tilde{\gamma}_c/\tilde{\gamma}_s$. 
This figure has been produced by numerically finding the roots of~\eqref{eq:sim_eqns_two_waves} for varying values of total sensing capacity $\tilde\chi$ and influx ratio $\tilde\gamma_c/\tilde\gamma_s$.
Rather than occupying diffuse regions of parameter space, the converged roots lie along narrow curved trajectories in each pairwise projection, indicating that the admissible states are strongly constrained by the coupled nonlinear balance equations. 
This demonstrates that the decay rates $(\lambda_F,\lambda_L)$ and concentrations $(\rho_c^{\dagger(F)},\rho_s^{\dagger(F)},\rho_s^{\dagger(L)})$ cannot vary independently, but instead evolve collectively along low-dimensional nonlinear manifolds determined by the interaction between chemotactic coupling and flux imbalance.
This highlights that heterogeneous collective migration cannot be inferred trivially from the behaviour of homogeneous populations, as the coupling between distinct subpopulations introduces additional nonlinear constraints and emergent dynamics that fundamentally alter the structure of the propagating fronts.

\section*{Discussion}\label{sec:conclusion}
Understanding the mechanisms shaping the co-migration of chemotactic cell populations is a fundamental problem in the study of collective cell dynamics, with relevance to processes ranging from bacterial colony expansion to cancer invasion and developmental patterning~\cite{Clark2015,McLennan2015,salek2019bacterial, Yamamoto2023,Qu2023}. 
In this work, we investigate the role of heterogeneity in the long-range migration of cell population via self-generated chemotaxis. Specifically, we extend the sensor/consumer framework proposed by~\cite{Ucar2025} to consider a binary cell mixture, where both cell types are capable of self-generated, sustained migration in response to a common chemoattractant. 
The model displays interesting travelling wave behaviour that we characterise by combining numerical and analytical results. 
As a starting point, we analyse the migration of homogeneous populations. 
In doing so, we extend the results of \cite{Ucar2025} by deriving an explicit expression for the travelling wave speed entirely in terms of the underlying model parameters. 
This analytical result provides direct insight into how migration dynamics depend on chemotactic sensitivity, consumption rates, and the rate at which cells enter the domain.

We then study the migration of heterogeneous consumer/sensor populations. 
Numerical simulations reveal that, at long time, the total cell density converges to a travelling terrace---a specific type of solution that consists of two coupled travelling waves that propagate at different speeds. 
To investigate the generality of these observations across parameter choices, we characterise travelling terrace solutions analytically. 
Specifically, we derive a set of nonlinear algebraic equations~\eqref{eq:sim_eqns_two_waves} defining the relationship between model parameters and the travelling speed of the two coupled travelling waves, and the spatial organisation of the different cell types in the different bulk regions. Although these equations cannot be solved explicitly, they still provide useful insights into the comigration pattern of consumer/sensor populations across parameter regimes.

Our propagating terrace analysis reveals the strong coupling between the two migrating fronts via the shaping of the chemoattractant profile. 
On the one hand, the leading population---consisting of sensor-only cells---modifies the chemoattractant landscape in a way that directly influences the behaviour of the following population---consisting of a mixture of sensor and consumer cells. At the same time, the following population simultaneously affects the bulk concentration, and, therefore, the propagation characteristics of the leading front. 
This mutual dependence demonstrates that the dynamics of heterogeneous migration cannot be understood by considering either population in isolation. Rather, the pattern of cell migration emerges from the collective interactions between cells mediated through the shared chemoattractant field, providing insight into the mechanisms required for stable and robust co-migration.

Our analytical results further show that the chemoattractant-mediated coupling has important consequences for the migration dynamics. 
Although the sensor and consumer homogeneous populations may possess identical intrinsic migratory speeds, introducing heterogeneity consistently breaks this symmetry. Specifically, consumer and sensor cells organise differently within the migrating front, producing distinct leading and following fronts with propagation speeds that deviate from the homogeneous benchmark. Using the derived analytical expression for the migration speeds of the two fronts, we investigate the extent to which heterogeneity impacts collective migration depends on the balance between sensing, chemoattractant consumption, and influx conditions (Figures~\ref{fig:figure3D}--\ref{fig:figure3E} and~\ref{fig:FIG_heterogeneous_param_dependence}). 

To allow for analytical progress, we made a number of simplifying assumptions that could be relaxed in future work. For example, our results both on homogeneous (\Cref{sec:homo}) and heterogeneous (\Cref{sec:heterogeneous chemotaxis}) self-generated chemotaxis rely on the chosen logarithmic form of the chemotactic response function~\eqref{eq:dim_sys}. 
While the logarithmic sensing law is well justified at moderate and high chemoattractant concentrations, cellular chemotactic sensitivity is expected to diminish below a critical concentration~\cite{bhattacharjee2021chemotactic,Ucar2025}. Incorporating this effect would alter the asymptotic behaviour of the model.
Specifically, we expect propagating terrace solutions to describe a transient rather than an asymptotic attractor for the migration dynamics. Properly understanding the behaviour of~\eqref{eq:dim_sys} under more general choices of the chemoattractant sensing function could therefore yield interesting mathematical challenges. Furthermore, we have neglected a range of mechanisms that could play a role in shaping collective migration of heterogenous cellular populations, such as mechanical interactions \cite{ArmstrongPainterSherratt,buttenschon2024cells,falco2022local, falco2025nonlocal, Ucar2025}. 
Since tissues often exhibit viscoelastic behaviour, mechanical interactions could provide an additional mechanism to maintain co-localisation between heterogenous subpopulations in dense environments \cite{Celora2026, Ford2025, Jewell2026}. 
Our model predicts that over time the two cell type separates, as the leading front which only consists of sensor cells separate over time at relative speed $\mathcal{V}_L-\mathcal{V}_F$. 
Incorporating mechanical adhesion or viscoelastic coupling may therefore stabilise the relative positions of the fronts and provide a more realistic description of heterogeneous collective migration in biological tissues. However, adhesion forces are likely insufficient to suppress the separation of the leading front due to their localised nature, as observed also in other migration models capture leader-follower asymmetry~\cite{Jewell2026}. Instead, preventing front separation may require longer-range mechanical or alignment interactions. Nonetheless, extending our analysis to incorporate mechanical interactions provides a natural step towards a models of heterogeneous collective migration of dense---rather than dispersed---cell population.

Overall, our work provides a comprehensive characterisation of chemotactic migration in heterogeneous sensor/ consumer cell populations and it reveals how the invasion dynamics emerge from the strong coupling and mutual dependence between two interacting fronts moving at different speeds.

\section*{Acknowledgments}
The authors thank Mehmet Can U\c{c}ar for helpful feedback and for providing the experimental images. For the purpose of Open Access, the authors have applied a CC BY public copyright licence to any Author Accepted Manuscript (AAM) version arising from this submission. CF and GLC acknowledge financial support from a Hooke Research Fellowship, and MW acknowledges funding from the UKRI-EPSRC (grant number EP/Y034791/1).

\appendix

\section{Non-dimensionalisation}\label{app:non_dim}
We here outline the non-dimensionalisation procedure used to obtain~\eqref{eq:nondim_sys}--\eqref{eq:nondim_BCs_rho} starting from~\eqref{eq:dim_sys}--\eqref{eq:BCs_dim}. In line with~\cite{Ucar2025}, we rescale space and time to balance the diffusion and dynamics of the chemoattractant by introducing the following scalings:
\begin{equation}\label{eq:scalings}
    x = \sqrt{\frac{D_a}{r \phi}} \bar x,\qquad t = \frac{1}{r \phi} \, \bar t, \qquad \rho_s = \
    \phi \bar\rho_s, \qquad \rho_c = \
    \phi \bar\rho_c, \qquad a = \alpha \bar a,
\end{equation}
where the bar is used to indicate non-dimensional variables. The positive constants $\phi$ and $\alpha$ are the reference cell and chemoattractant concentrations, respectively; here they are chosen to normalise the initial conditions (see~\Cref{app:ICs}). 
Substituting~\eqref{eq:scalings} into~\eqref{eq:dim_sys}--\eqref{eq:BCs_dim} and dropping the bars for clarity, we obtain the non-dimensional system:
\begin{subequations}
    \begin{align}
        \pdv{\rho_c}{t} &= \tilde D_c \pdv[2]{\rho_c}{x} - \kappa\tilde\chi \diffp{}{x} \left[ \rho_c \, \diffp{}{x} \Big[ \log(a) \Big]  \vphantom{\diffp{\rho}{x}} \right] ,\\
        \pdv{\rho_s}{t} &= \tilde D_s \pdv[2]{\rho_s}{x} - (1-\kappa)\tilde\chi \diffp{}{x} \left[ \rho_s \, \diffp{}{x} \Big[ \log(a) \Big]  \vphantom{\diffp{\rho}{x}} \right] ,\\
        \pdv{a}{t} &= \pdv[2]{a}{x} - \Big[ \omega \rho_s + (1-\omega)\rho_c \Big] a ,
    \end{align}
    coupled to the boundary conditions 
\begin{equation}
    \left. \pdv{a}{x} \right|_{x=0} = 0 \quad \text{and} \quad \lim_{x \to \infty} a = 1,
\end{equation}
and
\begin{equation}
    \tilde D_i \left. \pdv{\rho_i}{x} \right|_{x=0} = \tilde\gamma_i \quad \text{and} \quad \lim_{x \to \infty} \pdv{\rho_i}{x} = 0 ,
    \end{equation}\label{app:sys_non_dymensional}
\end{subequations}
In~\eqref{app:sys_non_dymensional}, we have introduced the following dimensionless constant \begin{equation}
\tilde\chi := \frac{\chi}{D_a}, \quad \tilde D_i := \frac{D_i}{D_a},\quad \tilde{\gamma}_i := \frac{ \gamma_i}{\sqrt{D_a r \phi^3}},\quad i=s,c.
\end{equation} 
Given an appropriate set of initial conditions (\Cref{app:ICs}), the dynamics of~\eqref{app:sys_non_dymensional} are therefore characterised by a total of 6 parameters (see~\Cref{tab:params}). Generally, we expect them to significantly vary depending on the biological system of interest. Here, for simplicity, we refer to the reference values from~\cite{Ucar2025} and investigate only the role that the parameters $\kappa$ and $\omega$ have in shaping the coupled invasion of sensor/consumer systems. 

\begin{table}[htb]
    \centering
        \caption{List of the dimensionless parameters in~\eqref{eq:nondim_sys}--\eqref{eq:nondim_BCs_rho}. Unless differently specified, the model parameters are set to the default value given, which corresponds to the one used in~\cite{Ucar2025} to describe the co-migration of dendritic and T cells.}
    \begin{tabular}{c||p{117mm} c c}
    \toprule
        Parameter & Description & Value(s) & Source\\
        \hline\addlinespace[2pt]
        $\tilde\chi$ & Ratio maximal cell chemotactic sensitivity to chemoattractant diffusivity &0.28&\cite{Ucar2025}\\
        $\tilde D_c$ & Ratio consumer cells' to chemoattractant diffusivities&0.07&\cite{Ucar2025}\\
        $\tilde D_s$ & Ratio sensor cells' sensitivity to chemoattractant diffusivities& 0.245&\cite{Ucar2025}\\
        $\tilde\gamma_c$ & Rescaled influx of consumer cells &0.05& --\\
        $\tilde\gamma_s$ & Rescaled influx of sensor cells &0.03& --\\
        $\kappa$ & Ratio of the consumer to sensor chemotactic sensitivity& (0,0.5)&--\\       $\omega$ & Ratio of the sensor to consumer chemoattractant consumption rate& (0,0.5)&--\\
        \bottomrule
    \end{tabular}
    \label{tab:params}
\end{table}
\section{Dynamic simulations of long-range migration}
\label{app:simulations}
In this section, we provide additional information regarding the numerical methods used to solve~\eqref{eq:nondim_sys}--\eqref{eq:nondim_BCs_rho} and the set-up of numerical simulations of long-rage cell migration.
\subsection{Initial conditions}\label{app:ICs}
To close~\eqref{eq:nondim_sys}--\eqref{eq:nondim_BCs_rho}, we are left to specify the initial distribution of consumer and sensor cells, as well as the initial chemoattractant profile. We note that, the long-term behaviour of the system---which is the focus of this work---is broadly independent of the choice of initial conditions (result not shown), provided they are consistent with the far-field conditions~\eqref{eq:nondim_BCs_a}--\eqref{eq:nondim_BCs_rho}. Hence, without loss of generality and in line with~\cite{Ucar2025}, we assume that the two populations are initially well-mixed and concentrated at the origin, while the chemoattractant distribution is homogeneous
\begin{equation}\label{eq:ICs_nondim}
    \rho_{s}(x) = \rho_{c}(x) = \frac{1}{1+e^{x}}, \quad a(x)\equiv 1.
\end{equation}
Equation~\eqref{eq:ICs_nondim} leads to exponential decay of the cell densities for $x\gg1$ and results in the solution to~\eqref{eq:nondim_sys}--\eqref{eq:nondim_BCs_rho} to quickly relax towards a travelling wave profile.

\subsection{Numerical Scheme}\label{app:finite_volume}
We solve~\eqref{eq:nondim_sys}--\eqref{eq:nondim_BCs_rho} numerically using the method of lines. 
Specifically, we discretise~\eqref{eq:nondim_sys}--\eqref{eq:nondim_BCs_rho} on the finite spatial domain $x\in[0,L]$, with $L\gg1$, using finite volumes following the work of Bailo et al.~\cite{nonlocalNumericalSchemeRafaMarkus}.
Although the model is posed on a semi-infinite domain with
\begin{equation}
    \lim_{x\to\infty}\rho_c=\lim_{x\to\infty}\rho_s=0,
\end{equation}
the numerical simulations are performed on the truncated spatial domain $[0,L]$. 
No-flux boundary conditions are imposed at the artificial boundary $x=L$, providing an accurate approximation of the asymptotic conditions provided that $L$ is chosen sufficiently large that the solution has decayed and its spatial gradients are negligible.

We start by rewriting Eqs.~\eqref{eq:nondim_sys_a} and \eqref{eq:nondim_sys_c} in the general form:
\begin{equation}\label{eq:general_conservation}
    \partial_t u + \pdv{}{x} \Big[ u V(\vec{u}) \Big] = r(\vec{u}),
\end{equation}
where $\vec{u}(x,t)=[\rho_c(x,t),\rho_s(x,t),a(x,t)]$, $V$ is the velocity vector, and $r$ is the reaction/source term. For the cell densities $\rho_c$ and $\rho_s$~\eqref{eq:nondim_sys_a}--\eqref{eq:nondim_sys_b}, the velocity $V$ is
\begin{equation}
    V:= \pdv{}{x} \Big( -\tilde D_i \log(\rho_i) + \tilde\chi_i \log(a) \Big) , \quad i=c,s,\label{app:Vrho}
\end{equation}
and the reaction term is $r\equiv0$. In the above, $\tilde\chi_c:=\kappa\tilde\chi$ and $\tilde\chi_s:=(1-\kappa)\tilde\chi$. For the chemoattractant~\eqref{eq:nondim_sys_c}, the velocity function is
\begin{equation}
    V:= \pdv{}{x} \left[ -\log(a) \right] ,\label{app:Va}
\end{equation}
and the reaction term is $r:= - (\omega\rho_s+(1-\omega)\rho_c) \, a$.

We divide the domain into $N$ cells $[x_{j}, x_{j+1}]$ of equal width $\Delta x = x_{j+1} - x_{j-1}$, and with centres $x_{j+1/2}= (j+1/2)\Delta x$ for $j=0,1,\ldots,N-1$. We set $x_0=0$ and $x_N=L$. 
We then approximate the solution at each cell centre $x_{j+1/2}$ as 
\begin{equation}
u(x_{j+1/2}, t) \approx \bar{u}_j(t) := \frac{1}{\Delta x} \int_{x_{j}}^{x_{j+1}} u(x, t) \, \dd x, \quad j=0,\ldots, N-1.
\end{equation}
Integrating Equation~\eqref{eq:general_conservation} over the $j$-th cell we obtain
\begin{equation}
\int_{x_{j}}^{x_{j+1}} \partial_t u \, \dd x + \Big[ u V(\vec{u}) \Big]_{x_{j}}^{x_{j+1}} = \int_{x_{j}}^{x_{j+1}} r\, \dd x.
\end{equation}
This rearranges to
\begin{equation}
\partial_t \bar{u}_j = -\frac{u_{j+1} V_{j+1} - u_{j} V_{j}}{\Delta x} + \bar{r}_j, \quad j=0,\ldots,N-1,\label{eq:discretise}
\end{equation}
where $u_j=u(x_j,t)$, $V_{j}:=V(\vec{u}_{j})$ is the velocity field evaluated at the cell edges and $\bar{r}_j$ is the cell-averaged source. We approximate the reaction term using the product of the cell-averaged quantities
\[\bar{r}_j:=\int_{x_{j}}^{x_{j+1}} r(\vec{u})\, \dd x\approx r(\bar{\vec{u}}_j), \quad j=0,\ldots,N-1,\]
while we construct the edge fluxes $u_{j}V_{j}$ via an upwind approximation
\begin{equation}
    u_{j} V_{j} \approx 
    \begin{cases} 
        \bar{u}_{j-1} V(u_{j}), & \text{if } V_{j} \geq 0, \\ 
        \bar{u}_{j} V(u_{j}), & \text{if } V_{j} < 0.
    \end{cases}
\end{equation} 
This choice ensures that information propagates in the physically correct direction and suppresses non-physical oscillations that can arise from using centred schemes in advection-dominated problems. Conveniently, $V$ can be written in the form $\pdv{}{x} \left[ v(\vec{u}) \right]$, both for cell densities~\eqref{app:Vrho} and the chemoattractant concentration~\eqref{app:Va}. We therefore approximate the interior fluxed using finite differences $$V_{j}\approx\frac{v_{j+1/2}-v_{j-1/2}}{\Delta x}, \quad j=1,\ldots, N-1,$$
where $v_{j+1/2}:=v(\bar{\vec{u}}_{j})$ can be directly estimated using the cell averages $\bar{\vec{u}}_j=[\rho_{c,j}(t),\rho_{s,j}(t),a_j(t)]$. At the end-point of the domain, the flux is instead imposed by the boundary conditions~\eqref{eq:nondim_BCs_a}--\eqref{eq:nondim_BCs_rho}. 
Finally, we solve the corresponding system of $3N$ coupled ordinary differential equations~\eqref{eq:discretise} using an adaptive explicit Runge--Kutta time-stepping algorithm, implemented in SciPy via \texttt{solve\_ivp} with the \texttt{RK45} method. The solution is sampled at $N_t$ equally spaced output times over the interval $[0,T]$, with a maximum internal step size of $0.05$.

\subsection{The role of Neumann condition at the inlet for the chemoattractant}\label{app:boundary_layer}
\begin{figure}[tb]
\begin{subfigure}{0\textwidth}
    \captionlistentry{}
    \label{Figure5A}
\end{subfigure}
\begin{subfigure}{0\textwidth}
    \captionlistentry{}
    \label{Figure5B}
\end{subfigure}
    \centering
    \includegraphics[width=\linewidth]{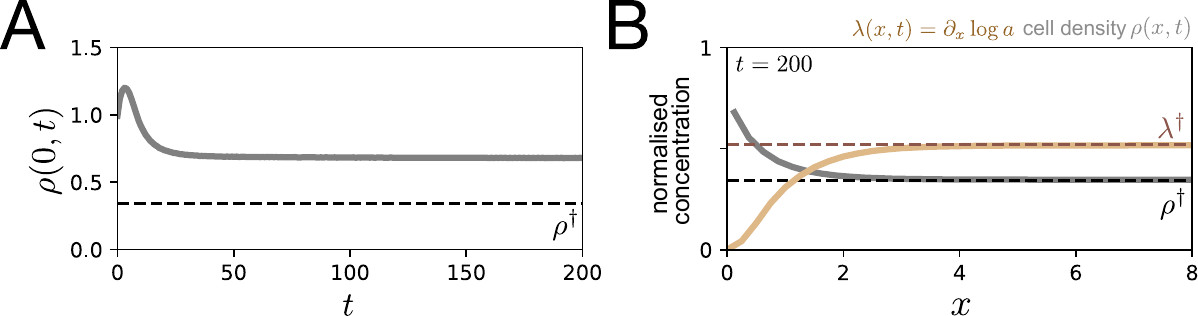}
    \caption{(A) Time evolution of cell concentration at the origin $\rho(0, t)$ for the same simulation as in~\Cref{fig:figure2A}. Eventually the concentration settles to a constant value which is higher than the bulk concentration $\rho^\dagger$~\eqref{eq:homo_analysic_solns}. (B) Asymptotic spatial distribution of the cell distribution $\rho$ and the chemotactic signal $\lambda=\partial_x\log a$ in the inlet boundary layer. Away from the inlet ($x=0$) the variables converge to corresponding bulk values ($\rho^\dagger$, $\lambda^\dagger$) defined by~\eqref{eq:homo_analysic_solns}.}
    \label{fig:rho_0_time}
\end{figure}

Following~\cite{Ucar2025}, we impose a no-flux boundary condition for the chemoattractant at the inlet of the domain (Figure~\ref{fig:hetero_schematic}).
This choice leads to the formation of a small boundary layer near the inlet ($x=0$)---see Figures~\ref{fig:figure2A} and~\ref{Figure5B}. In this region, the chemotactic signal $\partial_x\log(a)$ steeply increases from zero to its natural bulk value $\lambda^\dagger$ (\Cref{sec:TW homogeneous population}), while the cell density rapidly decreases to converge to the homogeneous bulk concentration. 
The accumulation of cell in the boundary layer is therefore due to the lack of directed cell motion---an artefact of the specific form of the boundary condition for the chemoattractant at $x=0$ and as such it is not of physical interest. The boundary layer disappears when changing the Neumann to the appropriate Robin conditions (results not shown)
$$\left.\partial_xa-\lambda a\right|_{x=0}=0.$$

Importantly, numerical simulations reveal that the solution profile in the boundary layer eventually converges to a stationary profile, while a travelling wave propagates from it (\Cref{Figure5A}). This is a key observation for the argument used in~\Cref{sec:TW homogeneous population} to derive a close form expression for the travelling wave speed. Specifically, the fact that the cell density eventually reaches a stationary distribution in the boundary layer implies that mass cannot be accumulate in this region of space and therefore the influx of cells at the inlet must be balanced by the outflux if cells in the bulk of the travelling wave.

\section{Additional results: propagating terrace analysis}\label{app:tw_analysis}
We here outline in more detailed the analytical steps to obtain~\eqref{eq:speed_leading}--\eqref{eq:sim_eq_1}.

We start by integrating Equations~\eqref{eq:front_2_ODEs_a}--\eqref{eq:front_2_ODEs_b} once. Imposing the far-field conditions at $\xi\to+\infty$~\eqref{eq:front_2_BSc_cells}, we obtain
\begin{subequations}\label{eq:front_2_app_ODEs}
    \begin{align}
        -\mathcal{V}_L P_c(\xi) &= \tilde D_c P_c'(\xi) - \kappa \tilde\chi P_c(\xi) \frac{\widetilde A'(\xi)}{\widetilde A(\xi)} ,\\
        -\mathcal{V}_L P_s(\xi) &= \tilde D_s P_s'(\xi) - (1-\kappa) \tilde\chi P_s(\xi) \frac{\widetilde A'(\xi)}{\widetilde A(\xi)}, \\
        -\mathcal{V}_L A'(\xi) &= A''(\xi) - ((1-\omega)P_c(\xi)+\omega P_s(\xi)) A(\xi) ,
     \end{align}
\end{subequations}
We compute the self-consistent behaviour of the chemoattractant field $A(\xi)$ at the back of the wave by linearising Equations~\eqref{eq:front_2_app_ODEs} around the steady state $(\rho_c^{\dagger(L)},\rho_s^{\dagger(L)},0)$ using the ansatz
\begin{equation}
(P_c,P_s,A)=(\rho_c^{\dagger(L)},\rho_s^{\dagger(L)},0)+\tilde{\vec{u}}e^{\lambda_L^\dagger \xi},\label{eq:linearisation leader} 
\end{equation}
where $\tilde{\vec{u}}$ is a constant vector with negligible magnitude $\|\vec{u}\|\ll 1$ capturing the direction of the unstable manifold and $\lambda_L^\dagger>0$ is the corresponding eigenvalue which characterises the rate at which the solution converges to $(\rho_c^{\dagger(L)},\rho_s^{\dagger(L)},0)$ as $z\to-\infty$. Substituting~\eqref{eq:linearisation leader} into~\eqref{eq:front_2_app_ODEs} and retaining only the leading order terms, we find the following algebraic system
\begin{subequations}
    \begin{align}
        \left(\mathcal{V}_L-\kappa\tilde{\chi}\lambda_L^\dagger\right) \rho_c^{\dagger (L)} &=0,\label{app:eqLrhoc}\\
       \left(\mathcal{V}_L-(1-\kappa)\tilde{\chi}\lambda_L^\dagger\right) \rho_s^{\dagger (L)} &= 0,\label{app:eqLrhos} \\
    (\lambda^\dagger_L)^2+\lambda^\dagger_L \mathcal{V}_L  - \Big[ \omega \rho_s^{\dagger (L)} + (1-\omega) \rho_c^{\dagger (L)}\Big]&=0 ,
     \end{align}\label{sys:front2_app_genera}
 \end{subequations}
for the bulk concentration $\rho_{c,s}^{\dagger(L)}$ and the decay rate $\lambda^\dagger_L$. Unless $\kappa=1/2$, then Equations~\eqref{app:eqLrhoc}--\eqref{app:eqLrhos} imply that either $\rho_s^{\dagger(L)}$ or $\rho_c^{\dagger(L)}$ has to vanish. Hence, the leader front is characterised by the segregation of one of the two populations. 

If we take $\rho_s^{\dagger(L)} \neq 0$,~\eqref{sys:front2_app_genera} reduces to the expression~\eqref{eq:speed_leading}--\eqref{eq:sim_eq_1}. In contrast, if we were to consider  $\rho_c^{\dagger(L)} \neq 0$, we would obtain the analogous conditions
\begin{equation}\label{eq:vel_lam_relation_front_2}
    \mathcal{V}_L = \kappa \tilde\chi \lambda_L,   \quad  \lambda_L^2(1 - \kappa \tilde\chi ) - (1-\omega) \rho_c^{\dagger(L)} = 0.
\end{equation}
While we do not observe this second scenario in the numerical simulations, at this stage, we have no reason to assume that it is not possible. The inconsistency of~\eqref{eq:vel_lam_relation_front_2} instead arise from looking at the behaviour in the follower front. Specifically integrating~\eqref{eq:front_1_ODEs_a}--\eqref{eq:front_1_ODEs_b} and imposing the far-field conditions~\eqref{eq:front_1_BCs_cells} and~\eqref{eq:vel_lam_relation_front_2}, we obtain
\begin{equation}
    \mathcal{V}_F=(1-\kappa)\tilde\chi\lambda_F,\label{app:VF}
\end{equation}
and
\begin{equation}
    (1-\kappa) \lambda_F \Big( \rho_c^{\dagger(L)} - \rho_c^{\dagger(F)} \Big) =  \kappa \Big( \rho_c^{\dagger(L)} \lambda_L - \rho_c^{\dagger(F)} \lambda_F \Big),
\end{equation}
which we can rearrange into
\begin{equation}
    \frac{\lambda_L\kappa}{(1-\kappa)\lambda_F}=1-(1-2\kappa)\frac{\rho_c^{\dagger(F)}}{\rho_c^{\dagger(L)}}.\label{app:ratio expression 1}
\end{equation}
Using~\eqref{app:VF} and~\eqref{eq:vel_lam_relation_front_2}, we can see that the left-hand side of~\eqref{app:ratio expression 1} is exactly the ratio between the leading and following fronts' travelling speeds
\begin{equation}
    \frac{\mathcal{V}_L}{\mathcal{V}_F}=1-(1-2\kappa)\frac{\rho_c^{\dagger(F)}}{\rho_c^{\dagger(L)}} ,
\end{equation}
which is less then $1$ since we assume $\kappa\in(0,0.5)$. We therefore find a contradiction, since the leading front, by definition, has to move faster. We conclude that the only possible propagating terrace solutions to Equations~\eqref{eq:nondim_sys} are characterised by a leading front that consists only of sensor cells while it excludes consumer cells.

\printbibliography

@article{Hillen2008,
  title = {{A user’s guide to PDE models for chemotaxis}},
  volume = {58},
  ISSN = {1432-1416},
  url = {http://dx.doi.org/10.1007/s00285-008-0201-3},
  DOI = {10.1007/s00285-008-0201-3},
  number = {1–2},
  journal = {Journal of Mathematical Biology},
  publisher = {Springer Science and Business Media LLC},
  author = {Hillen,  T. and Painter,  K. J.},
  year = {2008},
  pages = {183–217}
}

@article{Keller1971,
  title = {{Traveling bands of chemotactic bacteria: A theoretical analysis}},
  volume = {30},
  ISSN = {0022-5193},
  url = {http://dx.doi.org/10.1016/0022-5193(71)90051-8},
  DOI = {10.1016/0022-5193(71)90051-8},
  number = {2},
  journal = {Journal of Theoretical Biology},
  publisher = {Elsevier BV},
  author = {Keller,  Evelyn F. and Segel,  Lee A.},
  year = {1971},
  pages = {235–248}
}

@article{Ucar2025,
  title = {Self-generated chemotaxis of mixed cell populations},
  volume = {122},
  ISSN = {1091-6490},
  url = {http://dx.doi.org/10.1073/pnas.2504064122},
  DOI = {10.1073/pnas.2504064122},
  number = {34},
  journal = {Proceedings of the National Academy of Sciences},
  publisher = {Proceedings of the National Academy of Sciences},
  author = {U\c{c}ar,  M. and Alsberga,  Z. and Alanko,  J. and Sixt,  M. and Hannezo,  E.},
  year = {2025}
}

@article{Alert2022,
  title = {Cellular Sensing Governs the Stability of Chemotactic Fronts},
  volume = {128},
  ISSN = {1079-7114},
  url = {http://dx.doi.org/10.1103/PhysRevLett.128.148101},
  DOI = {10.1103/physrevlett.128.148101},
  number = {14},
  journal = {Physical Review Letters},
  publisher = {American Physical Society (APS)},
  author = {Alert,  Ricard and Martínez-Calvo,  Alejandro and Datta,  Sujit S.},
  year = {2022}
}

@article{Narla2021,
  title = {A traveling-wave solution for bacterial chemotaxis with growth},
  volume = {118},
  ISSN = {1091-6490},
  url = {http://dx.doi.org/10.1073/pnas.2105138118},
  DOI = {10.1073/pnas.2105138118},
  number = {48},
  journal = {Proceedings of the National Academy of Sciences},
  publisher = {Proceedings of the National Academy of Sciences},
  author = {Narla,  Avaneesh V. and Cremer,  Jonas and Hwa,  Terence},
  year = {2021}
}

@article{Mattingly2022,
  title = {Collective behavior and nongenetic inheritance allow bacterial populations to adapt to changing environments},
  volume = {119},
  ISSN = {1091-6490},
  url = {http://dx.doi.org/10.1073/pnas.2117377119},
  DOI = {10.1073/pnas.2117377119},
  number = {26},
  journal = {Proceedings of the National Academy of Sciences},
  publisher = {Proceedings of the National Academy of Sciences},
  author = {Mattingly,  Henry H. and Emonet,  Thierry},
  year = {2022}
}

@article{Vo2025,
  title = {Nongenetic adaptation by collective migration},
  volume = {122},
  ISSN = {1091-6490},
  url = {http://dx.doi.org/10.1073/pnas.2423774122},
  DOI = {10.1073/pnas.2423774122},
  number = {8},
  journal = {Proceedings of the National Academy of Sciences},
  publisher = {Proceedings of the National Academy of Sciences},
  author = {Vo,  Lam and Avgidis,  Fotios and Mattingly,  Henry H. and Edmonds,  Karah and Burger,  Isabel and Balasubramanian,  Ravi and Shimizu,  Thomas S. and Kazmierczak,  Barbara I. and Emonet,  Thierry},
  year = {2025}
}

@article{stock2022self,
  title={{A self-generated Toddler gradient guides mesodermal cell migration}},
  author={Stock, Jessica and Kazmar, Tomas and Schlumm, Friederike and Hannezo, Edouard and Pauli, Andrea},
  journal={Science Advances},
  volume={8},
  number={37},
  pages={eadd2488},
  year={2022},
  doi={10.1126/sciadv.add2488},
  publisher={American Association for the Advancement of Science}
}

@article{dona2013directional,
  title={Directional tissue migration through a self-generated chemokine gradient},
  author={Don{\`a}, Erika and Barry, Joseph D and Valentin, Guillaume and Quirin, Charlotte and Khmelinskii, Anton and Kunze, Andreas and Durdu, Sevi and Newton, Lionel R and Fernandez-Minan, Ana and Huber, Wolfgang and others},
  journal={Nature},
  volume={503},
  number={7475},
  pages={285--289},
  year={2013},
  doi={10.1038/nature12635},
  publisher={Nature Publishing Group UK London}
}

@article{salek2019bacterial,
  title={{Bacterial chemotaxis in a microfluidic T-maze reveals strong phenotypic heterogeneity in chemotactic sensitivity}},
  author={Salek, M Mehdi and Carrara, Francesco and Fernandez, Vicente and Guasto, Jeffrey S and Stocker, Roman},
  journal={Nature Communications},
  volume={10},
  number={1},
  pages={1877},
  year={2019},
  doi={10.1038/s41467-019-09521-2},
  publisher={Nature Publishing Group UK London}
}

@article{p2025phenotypic,
  title={Phenotypic heterogeneity in temporally fluctuating environments},
  author={P Browning, Alexander and Hamis, Sara},
  journal={Physical Biology},
  volume={22},
  number={5},
  pages={056002},
  year={2025},
  doi={10.1088/1478-3975/adf790},
  publisher={IOP Publishing}
}

@article{bhattacharjee2021chemotactic,
  title={Chemotactic migration of bacteria in porous media},
  author={Bhattacharjee, Tapomoy and Amchin, Daniel B and Ott, Jenna A and Kratz, Felix and Datta, Sujit S},
  journal={Biophysical Journal},
  volume={120},
  number={16},
  pages={3483--3497},
  year={2021},
  doi={10.1016/j.bpj.2021.05.012},
  publisher={Elsevier}
}

@article{blanchet2008infinite,
  title={{Infinite time aggregation for the critical Patlak-Keller-Segel model in $\mathbb{R}^2$}},
  author={Blanchet, Adrien and Carrillo, Jos{\'e} A and Masmoudi, Nader},
  journal={Communications on Pure and Applied Mathematics},
  volume={61},
  number={10},
  pages={1449--1481},
  year={2008},
  doi={10.1002/cpa.20225},
  publisher={Wiley Online Library}
}

@article{calvez2006volume,
  title={{Volume effects in the Keller--Segel model: energy estimates preventing blow-up}},
  author={Calvez, Vincent and Carrillo, Jos{\'e} A},
  journal={Journal de Math{\'e}matiques Pures et Appliqu{\'e}es},
  volume={86},
  number={2},
  pages={155--175},
  year={2006},
  doi={10.1016/j.matpur.2006.04.002},
  publisher={Elsevier}
}

@article{CalvezCorrias2008,
  author       = {Calvez, Vincent and Corrias, Lucilla},
  title        = {The parabolic-parabolic {Keller--Segel} model in {$\mathbb{R}^2$}},
  journal      = {Communications in Mathematical Sciences},
  year         = {2008},
  volume       = {6},
  number       = {2},
  doi={10.4310/CMS.2008.v6.n2.a8},
  pages        = {417--447}
}

@article{Ducrot2012ExistenceAC,
  title={Existence and convergence to a propagating terrace in one-dimensional reaction-diffusion equations},
  author={Arnaud Ducrot and Thomas Giletti and Hiroshi Matano},
  journal={Transactions of the American Mathematical Society},
  year={2012},
  volume={366},
  pages={5541-5566},
  doi={10.1090/S0002-9947-2014-06105-9},
  url={https://api.semanticscholar.org/CorpusID:10855528}
}

@article{SaragostiEtAl2010,
  author       = {Saragosti, Jonathan and Calvez, Vincent and Bournaveas, Nikolaos and Buguin, Axel and Silberzan, Pascal and Perthame, Beno\^{\i}t},
  title        = {Mathematical Description of Bacterial Traveling Pulses},
  journal      = {PLOS Computational Biology},
  year         = {2010},
  volume       = {6},
  number       = {8},
  pages        = {e1000890},
  doi          = {10.1371/journal.pcbi.1000890}
}

@article{nonlocalNumericalSchemeRafaMarkus,
author = {Bailo, Rafael and Carrillo, J. A. and Murakawa, Hideki and Schmidtchen, Markus},
year = {2020},
title = {{Convergence of a fully discrete and energy-dissipating finite-volume scheme for aggregation-diffusion equations}},
volume = {30},
number = {13},
pages = {2487--2522},
journal = {Mathematical Models and Methods in Applied Sciences},
doi = {10.1142/S0218202520500487}
}

@article{carrillo2015finite,
  title   = {A Finite-Volume Method for Nonlinear Nonlocal Equations with a Gradient Flow Structure},
  author  = {Carrillo, Jos{\'e} A. and Chertock, Alina and Huang, Yanghong},
  journal = {Communications in Computational Physics},
  volume  = {17},
  number  = {1},
  pages   = {233--258},
  year    = {2015},
  doi     = {10.4208/cicp.160214.010814a}
}

@article{Ford2025,
  title = {Pattern formation along signaling gradients driven by active droplet behavior of cell swarms},
  volume = {122},
  ISSN = {1091-6490},
  url = {http://dx.doi.org/10.1073/pnas.2419152122},
  DOI = {10.1073/pnas.2419152122},
  number = {21},
  journal = {Proceedings of the National Academy of Sciences},
  publisher = {Proceedings of the National Academy of Sciences},
  author = {Ford,  Hugh Z. and Celora,  Giulia L. and Westbrook,  Elizabeth R. and Dalwadi,  Mohit P. and Walker,  Benjamin J. and Baumann,  Hella and Weijer,  Cornelis J. and Pearce,  Philip and Chubb,  Jonathan R.},
  year = {2025},
}

@misc{Celora2026,
  doi = {10.48550/ARXIV.2602.20088},
  url = {https://arxiv.org/abs/2602.20088},
  author = {Celora,  Giulia L. and Walker,  Benjamin J. and Dalwadi,  Mohit P. and Pearce,  Philip},
  title = {Chemotaxis of cell aggregates: morphology and dynamics of migrating active droplets},
  eprint = {2602.20088},
  publisher = {arXiv},
  year = {2026},
  archivePrefix = {arXiv},
  copyright = {Creative Commons Attribution 4.0 International}
}

@article{falco2022local,
  author       = {Falc{\'o}, C. and Baker, R.~E. and Carrillo, J.~A.},
  title        = {{A local continuum model of cell–cell adhesion}},
  journal      = {SIAM Journal on Applied Mathematics},
  volume       = {84},
  number       = {3},
  pages        = {S17--S42},
  doi={10.1137/22M1506079},
  year         = {2024}
}

@article{falco2025nonlocal,
    author = {Falc\'{o}, C. and Baker, R. E. and Carrillo, J. A.},
    title = {A nonlocal-to-local approach to aggregation-diffusion equations},
    journal = {SIAM Review},
    volume = {67},
    number = {2},
    pages = {353-372},
    doi={10.1137/25M1726248},
    year = {2025}
}

@article{buttenschon2024cells,
  author       = {Buttensch{\"o}n, A. and Sinclair, S. and Edelstein-Keshet, L.},
  title        = {{How cells stay together: a mechanism for maintenance of a robust cluster explored by local and non-local continuum models}},
  journal      = {Bulletin of Mathematical Biology},
  volume       = {86},
  number       = {11},
  pages        = {129},
  doi={10.1007/s11538-024-01355-4},
  year         = {2024}
}

@article{ArmstrongPainterSherratt,
title = {A continuum approach to modelling cell–cell adhesion},
journal = {{Journal of Theoretical Biology}},
volume = {243},
number = {1},
pages = {98-113},
year = {2006},
doi = {10.1016/j.jtbi.2006.05.030},
author = {Nicola J. Armstrong and Kevin J. Painter and Jonathan A. Sherratt}
}

@misc{Jewell2026,
  doi = {10.48550/arXiv.2604.15283},
  url = {https://arxiv.org/abs/2604.15283},
  author = {Jewell,  Thomas Jun and Johnson,  Samuel W. S. and Baker,  Ruth E. and Maini,  Philip K.},
  title = {Cell-cell adhesion cannot sustain extended follower streams in a minimal non-local model of leader-follower migration},
  archivePrefix = {arXiv},
  eprint = {2604.15283},
  publisher = {arXiv},
  year = {2026},
  copyright = {Creative Commons Attribution 4.0 International}
}

@book{Phillips2025,
  author    = {Ben Phillips},
  title     = {The Ecology and Evolution of Invasive Populations},
  chapter   = {Pushed and Pulled Waves},
  publisher = {Oxford Academic},
  year      = {2025}
}

@article{McLennan2015,
  title = {Neural crest migration is driven by a few trailblazer cells with a unique molecular signature narrowly confined to the invasive front},
  volume = {142},
  ISSN = {0950-1991},
  url = {http://dx.doi.org/10.1242/dev.117507},
  DOI = {10.1242/dev.117507},
  number = {11},
  journal = {Development},
  publisher = {The Company of Biologists},
  author = {McLennan,  Rebecca and Schumacher,  Linus J. and Morrison,  Jason A. and Teddy,  Jessica M. and Ridenour,  Dennis A. and Box,  Andrew C. and Semerad,  Craig L. and Li,  Hua and McDowell,  William and Kay,  David and Maini,  Philip K. and Baker,  Ruth E. and Kulesa,  Paul M.},
  year = {2015},
  pages = {2014–2025}
}

@article{Ratnayake2021,
  title = {Macrophages provide a transient muscle stem cell niche via NAMPT secretion},
  volume = {591},
  ISSN = {1476-4687},
  url = {http://dx.doi.org/10.1038/s41586-021-03199-7},
  DOI = {10.1038/s41586-021-03199-7},
  number = {7849},
  journal = {Nature},
  publisher = {Springer Science and Business Media LLC},
  author = {Ratnayake,  Dhanushika and Nguyen,  Phong D. and Rossello,  Fernando J. and Wimmer,  Verena C. and Tan,  Jean L. and Galvis,  Laura A. and Julier,  Ziad and Wood,  Alasdair J. and Boudier,  Thomas and Isiaku,  Abdulsalam I. and Berger,  Silke and Oorschot,  Viola and Sonntag,  Carmen and Rogers,  Kelly L. and Marcelle,  Christophe and Lieschke,  Graham J. and Martino,  Mikaël M. and Bakkers,  Jeroen and Currie,  Peter D.},
  year = {2021},
  pages = {281–287}
}

@article{Alanko2023,
  title = {CCR7 acts as both a sensor and a sink for CCL19 to coordinate collective leukocyte migration},
  volume = {8},
  ISSN = {2470-9468},
  url = {http://dx.doi.org/10.1126/sciimmunol.adc9584},
  DOI = {10.1126/sciimmunol.adc9584},
  number = {87},
  journal = {Science Immunology},
  publisher = {American Association for the Advancement of Science (AAAS)},
  author = {Alanko,  Jonna and U\c{c}ar,  Mehmet Can and Canigova,  Nikola and Stopp,  Julian and Schwarz,  Jan and Merrin,  Jack and Hannezo,  Edouard and Sixt,  Michael},
  year = {2023},
}

@article{Clark2015,
  title = {Modes of cancer cell invasion and the role of the microenvironment},
  volume = {36},
  ISSN = {0955-0674},
  url = {http://dx.doi.org/10.1016/j.ceb.2015.06.004},
  DOI = {10.1016/j.ceb.2015.06.004},
  journal = {Current Opinion in Cell Biology},
  publisher = {Elsevier BV},
  author = {Clark,  Andrew G and Vignjevic,  Danijela Matic},
  year = {2015},
  pages = {13–22}
}

@article{Qu2023,
  title = {Crosstalk between small-cell lung cancer cells and astrocytes mimics brain development to promote brain metastasis},
  volume = {25},
  ISSN = {1476-4679},
  url = {http://dx.doi.org/10.1038/s41556-023-01241-6},
  DOI = {10.1038/s41556-023-01241-6},
  number = {10},
  journal = {Nature Cell Biology},
  publisher = {Springer Science and Business Media LLC},
  author = {Qu,  Fangfei and Brough,  Siqi C. and Michno,  Wojciech and Madubata,  Chioma J. and Hartmann,  Griffin G. and Puno,  Alyssa and Drainas,  Alexandros P. and Bhattacharya,  Debadrita and Tomasich,  Erwin and Lee,  Myung Chang and Yang,  Dian and Kim,  Jun and Peiris-Pagès,  Maria and Simpson,  Kathryn L. and Dive,  Caroline and Preusser,  Matthias and Toland,  Angus and Kong,  Christina and Das,  Millie and Winslow,  Monte M. and Pasca,  Anca M. and Sage,  Julien},
  year = {2023},
  pages = {1506–1519}
}

@article{Yamamoto2023,
  title = {Orchestration of Collective Migration and Metastasis by Tumor Cell Clusters},
  volume = {18},
  ISSN = {1553-4014},
  url = {http://dx.doi.org/10.1146/annurev-pathmechdis-031521-023557},
  DOI = {10.1146/annurev-pathmechdis-031521-023557},
  number = {1},
  journal = {Annual Review of Pathology: Mechanisms of Disease},
  publisher = {Annual Reviews},
  author = {Yamamoto,  Ami and Doak,  Andrea E. and Cheung,  Kevin J.},
  year = {2023},
  pages = {231–256}
}

@book{murray_biological_2002,
  	author = {Murray, J. D.},
	title = {Biological Waves: Single-Species Models},
	publisher = {Springer},
	year = {2002},
	pages = {437--483},
}

@article{simpson_fisherkpp-type_2024,
	title = {Fisher–{KPP}-type models of biological invasion: open source computational tools, key concepts and analysis},
	volume = {480},
	issn = {1364-5021},
	url = {https://doi.org/10.1098/rspa.2024.0186},
	doi = {10.1098/rspa.2024.0186},
	number = {2294},
	journal = {Proceedings of the Royal Society A: Mathematical, Physical and Engineering Sciences},
	author = {Simpson, Matthew J. and McCue, Scott W.},
	year = {2024},
	pages = {20240186},
}

@article{crossley_phenotypic_2024,
	title = {Phenotypic switching mechanisms determine the structure of cell migration into extracellular matrix under the ‘go-or-grow’ hypothesis},
	volume = {374},
	issn = {0025-5564},
	url = {https://www.sciencedirect.com/science/article/pii/S0025556424001007},
	doi = {10.1016/j.mbs.2024.109240},
	urldate = {2026-07-07},
	journal = {Mathematical Biosciences},
	author = {Crossley, Rebecca M. and Painter, Kevin J. and Lorenzi, Tommaso and Maini, Philip K. and Baker, Ruth E.},
	year = {2024},
	pages = {109240},
}

@article{macfarlane_impact_2022,
	title = {The Impact of Phenotypic Heterogeneity on Chemotactic Self-Organisation},
	volume = {84},
	issn = {1522-9602},
	doi = {10.1007/s11538-022-01099-z},
	language = {en},
	number = {12},
	journal = {Bulletin of Mathematical Biology},
	author = {Macfarlane, Fiona R. and Lorenzi, Tommaso and Painter, Kevin J.},
	year = {2022},
	pages = {143},
}

@article{ferguson_statistical_2017,
	title = {Statistical {Inference} of {The} {Mechanisms} {Driving} {Collective} {Cell} {Movement}},
	volume = {66},
	issn = {0035-9254},
	url = {https://doi.org/10.1111/rssc.12203},
	doi = {10.1111/rssc.12203},
	number = {4},
	journal = {Journal of the Royal Statistical Society Series C: Applied Statistics},
	author = {Ferguson, Elaine A. and Matthiopoulos, Jason and Insall, Robert H. and Husmeier, Dirk},
	year = {2017},
	pages = {869--890},
}

@misc{Celora2026-2,
  doi = {10.48550/arXiv.2606.17891},
  url = {https://arxiv.org/abs/2606.17891},
  author = {Celora,  Giulia L. and Watts,  Marjorie and Falcó,  Carles and Dalwadi,  Mohit P.},
  title = {A nonlinear theory for chemotactic fronts of mixed populations},
  eprint = {2606.17891},
  publisher = {arXiv},
  year = {2026},
  archivePrefix = {arXiv},
  copyright = {Creative Commons Attribution 4.0 International}
}

@article{lorenzi_phenotype_2025,
	title = {Phenotype structuring in collective cell migration: a tutorial of mathematical models and methods},
	volume = {90},
	issn = {1432-1416},
	shorttitle = {Phenotype structuring in collective cell migration},
	doi = {10.1007/s00285-025-02223-y},
	number = {6},
	urldate = {2026-03-23},
	journal = {Journal of Mathematical Biology},
	author = {Lorenzi, Tommaso and Painter, Kevin J. and Villa, Chiara},
	year = {2025},
	pages = {61}
}

@article{freingruber_trait-structured_2025,
	title = {Trait-structured chemotaxis: exploring ligand-receptor dynamics and travelling wave properties in a {Keller}–{Segel} model},
	volume = {38},
	shorttitle = {Trait-structured chemotaxis},
	doi = {10.1088/1361-6544/ae03ff},
	number = {10},
	urldate = {2026-03-23},
	journal = {Nonlinearity},
	author = {Freingruber, Viktoria and Lorenzi, Tommaso and Painter, Kevin J and Ptashnyk, Mariya},
	year = {2025},
	pages = {105006}
}

\end{document}